\documentclass[aps,twocolumn,nofootinbib,preprintnumbers,superscriptaddress,eqsecnum]{revtex4-1}

\usepackage{color}
\usepackage[
      colorlinks=true,
      linkcolor=blue,
      urlcolor=blue,
      filecolor=black,
      citecolor=red,
      pdfstartview=FitV,
      pdftitle={},
        pdfauthor={},
        pdfsubject={},
        pdfkeywords={},
        pdfpagemode=None,
        bookmarksopen=true
      ]{hyperref}

\newcommand{\HL}[1]{\textcolor{black}{#1}}

\usepackage[normalem]{ulem}
\usepackage{amsmath}
\usepackage{enumerate}
\usepackage{amsfonts}
\usepackage{epsfig}

\newcommand{\be}{\begin{equation}}
\newcommand{\ee}{\end{equation}}
\newcommand{\bea}{\begin{eqnarray}}
\newcommand{\eea}{\end{eqnarray}}

\newcommand{\bln}{\begin{align}}
\newcommand{\eln}{\end{align}}
\newcommand{\bst}{\begin{split}}
\newcommand{\est}{\end{split}}
\newcommand{\bi}{\begin{itemize}}
\newcommand{\ei}{\end{itemize}}
\newcommand{\ben}{\begin{enumerate}}
\newcommand{\een}{\end{enumerate}}

\def\ov{\over}
\def\le{\left}
\def\ri{\right}
\def\ha{{1\over 2}}
\def\lam{{\lambda}}
\def\Lam{{\Lambda}}
\def\Sig{{\Sigma}}
\def\al{{\alpha}}

\def\det{{\rm det}}
\def\tr{{\rm tr}}
\def\Tr{{\rm Tr}}

\def\th{{\theta}}

\def\Om{{\Omega}}
\def \th{{\theta}}

\def \lam {\lambda}
\def \om {\omega}

\def\sig{{\sigma}}
\def\ep{{\epsilon}}

\newcommand{\p}{\partial}

\newcommand\Ga{{\ensuremath{{\Gamma}}}}

\def\lam{{\lambda}}

\def\eeq{\end{equation}}

\def\Im{\mathop{\rm Im}}

\def\Tr{\mathop{\rm Tr}}

\newcommand\sA{{\ensuremath{{\mathcal A}}}}

\newcommand\sD{{\ensuremath{{\mathcal D}}}}

\newcommand\sN{{\ensuremath{{\mathcal N}}}}
\newcommand\sO{{\ensuremath{{\mathcal O}}}}

\newcommand\sJ{{\mathcal J}}

\newcommand\jT{\rho}

\newcommand\bpsi{{\bar \psi}}
\newcommand\bPsi{{\overline \Psi}}
\newcommand\bPhi{{\overline \Phi}}

\newcommand\ur{{\underline{r}}}

\newcommand\umu{{\underline{\mu}}}

\newcommand\utau{{\underline{\tau}}}

\begin{document}

\title {Luttinger's theorem, superfluid vortices, and holography}

\preprint{NSF-KITP-11-265, MIT-CTP 4333}

\author{Nabil Iqbal}
\affiliation{Kavli Institute for Theoretical Physics, University of California,
Santa Barbara, CA 93106}

\author{Hong Liu}

\affiliation{Center for Theoretical Physics,
Massachusetts
Institute of Technology,
Cambridge, MA 02139
}


\begin{abstract}

Strongly coupled field theories with gravity duals can be placed at finite density in two ways: electric field flux emanating from behind a horizon, or bulk charged fields outside of the horizon that explicitly source the density. We discuss field-theoretical observables that are sensitive to this distinction. If the charged fields are fermionic, we discuss a modified Luttinger's theorem that holds for holographic systems, in which the sum of boundary theory Fermi surfaces counts only the charge outside of the horizon. If the charged fields are bosonic, we show that the the resulting superfluid phase may be characterized by the coefficient of the transverse Magnus force on a moving superfluid vortex, which again is sensitive only to the charge outside of the horizon. For holographic systems these observables provide a field-theoretical way to distinguish how much charge is held by a dual horizon, but they may be useful in more general contexts as measures of deconfined (i.e. ``fractionalized'') charge degrees of freedom. 

\end{abstract}

\maketitle

\tableofcontents

\section{Introduction}\label{intro}

This paper will be concerned with holographic descriptions of compressible phases of matter. Recall that a compressible phase is conveniently defined to be one in which a $U(1)$ charge density $\jT $ is a continuous function of various parameters such as the chemical potential $\mu$ and the temperature $T$. When the $U(1)$ symmetry is unbroken, as emphasized recently in \cite{Huijse:2011hp}, all known field-theoretical examples  invariably involve Fermi surfaces that carry the $U(1)$ charge, and one expects Luttinger's theorem, which relates the charge density to the volume enclosed by Fermi surfaces, to apply.  

On the other hand, if the $U(1)$ symmetry is broken, the system is in a superfluid state. While Luttinger's theorem no longer applies, there nevertheless does exist a probe of the charge density, which is provided by a vortex. If one moves a vortex in a closed loop the many-body wavefunction acquires a Berry phase; relatedly, a moving vortex in a superfluid feels a transverse force called the Magnus force. As we explain later, this Berry phase and Magnus force can be viewed as probes of the charge density. 

The goal of this paper is to examine the status of these relations in holographic systems. 
We find that in both situations (whether the global $U(1)$ is broken or not), these relations develop anomalous contributions which are associated with horizon degrees of freedom in the bulk, and which can in turn be interpreted in the boundary theory as deconfined (or ``fractionalized'') degrees of freedom with respect to the gauge group of the boundary theory. 

\subsection{Luttinger deficit}

A fundamental  result of Fermi liquid theory is Luttinger's theorem \cite{Luttinger}, which in its original form states that to all orders in perturbation theory  the volume $V$ enclosed in a Fermi surface of electron is equal to the charge density $\rho$:
\be
\frac{e}{(2\pi)^{d-1}}  V = \jT \label{lutt} \ .
\ee
The original proof of Luttinger and Ward was based on perturbation theory. More recently an elegant nonperturbative proof was given by Oshikawa in~\cite{Oshikawa}, which highlighted the topological nature of the theorem. Oshikawa's proof also admits a generalization to exotic phases of matter in which the electron fractionalizes into constituent particles with different quantum numbers. Such a fractionalization is accompanied by an emergent gauge symmetry. It was found that in such a case Luttinger's theorem can be violated, with the deficit being related to the momentum carried by a topological excitation of the emergent gauge theory~\cite{FFL,Z2frac, Paramekanti2004}. 
 
We now turn to holography. The field theories which are known to admit gravity duals all involve non-Abelian gauge theories with various matter degrees of freedom \cite{AdS/CFT}. In such a theory we will define 
the location of a Fermi surface as a shell in momentum space with a radius $k_F$ which satisfies 
\be \label{defFr}
G^{-1} (\om=0, k_F) = 0 \ 
\ee
where $G$ is the retarded Green function of any {\it gauge invariant} fermionic operator. The generalization of the Luttinger theorem~\eqref{lutt} to such theories would then naively amount to 
to
\be
\frac{1}{(2\pi)^{d-1}}\sum_i q_i V_i = \jT  \label{lutorig} \ 
\ee
where the sum is over all Fermi surfaces of charged fermionic operators of charge $q_i$. 

Now recall that the charge density $\jT$ of the boundary theory is mapped to the boundary value of the radial {\it electric field} in the gravitational bulk, i.e. the gauge-gravitational dictionary gives us
\be
\jT = \frac{1}{g_F^2} \sqrt{-g}F^{tr}(r \to \infty) \ . \label{rhodef}
\ee
Here $g_F$ is a bulk gauge coupling and $F^{tr}$ is the radial electric field component of the bulk $U(1)$ gauge field corresponding the boundary current. Thus if we are studying a field theory state 
with a nonzero $\rho$, something must create a nonzero electric field in the bulk. There are two known ways to implement this: 

\ben
\item One may study a system with a charged black hole horizon in the bulk. In this case the source of the electric field is actually hidden behind the horizon. Nowhere explicitly in the bulk geometry does one actually see any charged matter, and the electric field flux is thus constant throughout the geometry. One example of this is the Reissner-Nordstrom black hole \cite{RNBH}, which is characterized by its AdS$_2 \times \mathbb{R}^{d-2}$ infrared geometry at zero temperatures. A different example is provided by the dilatonic systems studied in \cite{Goldstein:2009cv,Charmousis:2010zz,Gubser:2009qt}, which also have degenerate horizons that in the infrared source the flux. 

\item One may explicitly introduce bulk charged fields that source the electric field. If these charge carriers are fermionic then the eventual ground state of the system generally involves a gas of fermions filling (much of) the bulk geometry \cite{Hartnoll:2009ns,Hartnoll:2010gu} (\HL{see also~\cite{Arsiwalla:2010bt,vCubrovic:2010bf,Harada:2011aa}}): this state was dubbed the ``electron star'' by \cite{Hartnoll:2010gu}. If these charge carriers are bosonic then at sufficiently low temperatures the scalar generally condenses in the bulk, breaking the $U(1)$ symmetry: this is the well-studied holographic superfluid \cite{holographicsc}, and will be discussed in the next section. 
\een
It is of course possible to consider these two situations coexisting; i.e. there could be a horizon with a nonzero electric field, together with some charged matter outside the horizon that also contributes to the final boundary value of $F^{rt}$. Indeed, at finite temperature this is generically the case, and the recent construction of \cite{arXiv:1111.2606} realizes this explicitly at zero temperature.

\begin{figure}[h]
\begin{center}
\includegraphics[scale=0.40]{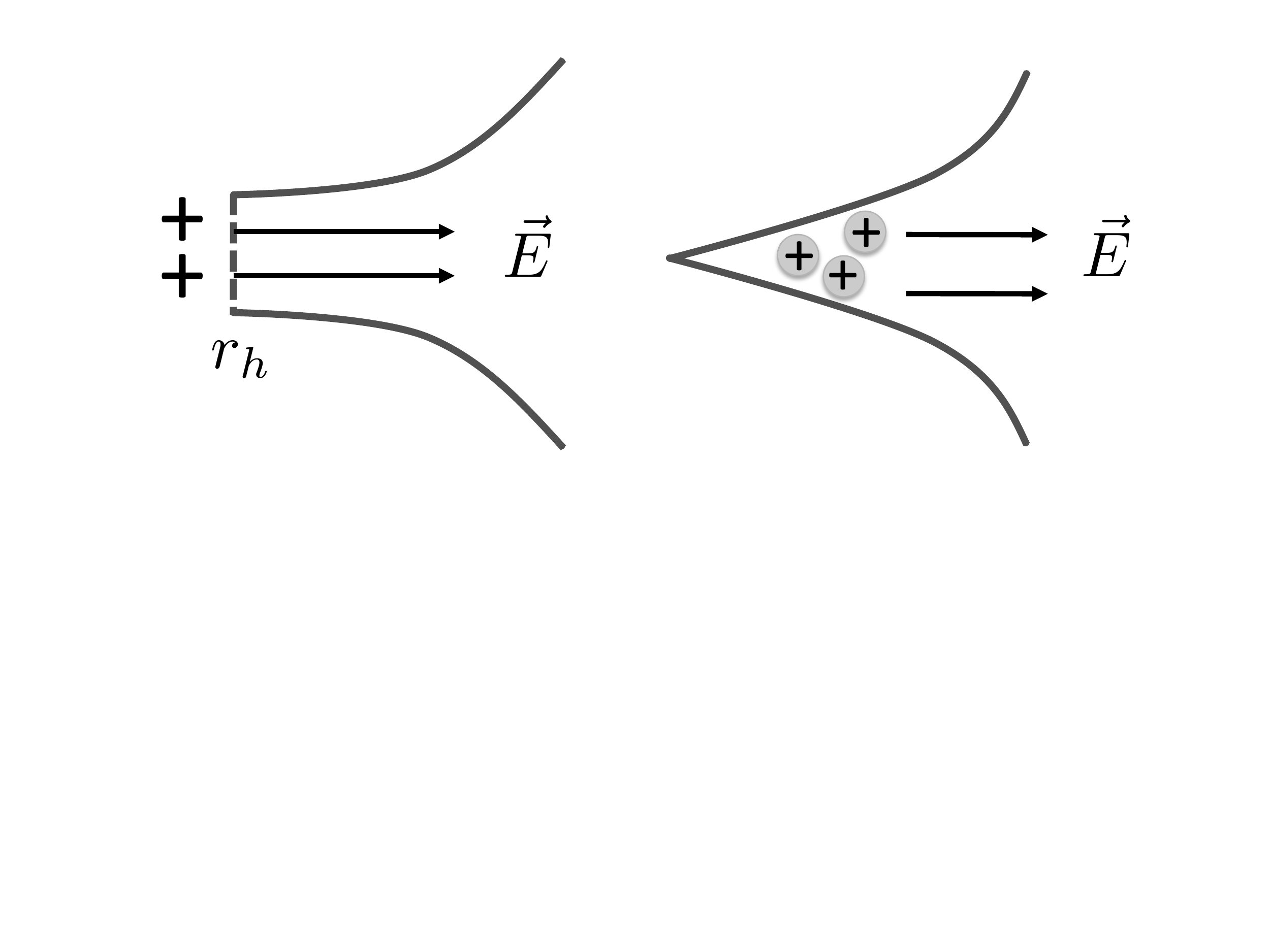}
\end{center}
\vskip -0.5cm
\caption{Different ways to create a finite $U(1)$ density in a field theory with a holographic description. {\it Left:} The electric field at infinity is sourced by a horizon. {\it Right:} The electric field is sourced by charged fields in the bulk.}
\label{fig:spacetimes}
\end{figure}

We will show that for a holographic system in the above setup,  instead of~\eqref{lutorig} one has  
\be
\frac{1}{(2\pi)^{d-1}}\sum_i q_i V_i = \jT - \sA \label{lutform}
\ee
where the deficit $\sA$ being given by the electric field flux through the horizon of the bulk geometry, i.e. 
given by 
\be
\sA = \le(-\frac{1}{g_F^2}\sqrt{-g}F^{rt}(r_h)\ri) \ .\label{adef1}
\ee
Equation~\eqref{lutform} means that the total Fermi surface volume only counts the charge density {\it outside} the horizon. 

This formula is not  novel. When $A=0$, it was previously shown in specific examples in ~\cite{Hartnoll:2011dm,Iqbal:2011in}.
In the form written above the formula was conjectured in \cite{arXiv:1106.4324} and also follows from the discussion in \cite{arXiv:1006.3794,arXiv:1012.0299,Sachdev:2011ze}. In particular an elegant and general proof for a holographic system with no horizon was given in \cite{Sachdev:2011ze}. Our derivation is along similar lines as \cite{Sachdev:2011ze}, but differs in some details. We work in a more general situation, allowing for a horizon and the associated nontrivial widths of fermionic excitations.


We now discuss the standard understanding of such a deficit, as articulated previously by various authors \cite{Huijse:2011hp,Iqbal:2011in,arXiv:1106.4324,SachdevNew,arXiv:1006.3794,arXiv:1012.0299}. The gapless excitations which characterize a Fermi surface (as defined in~\eqref{defFr}) are excitations of gauge invariant operators, and may be considered ``confined'' 
degrees of freedom, corresponding to the quanta of mesonic bound states in the gauge theory. On the other hand, in holography \cite{Witten:1998zw}, one thinks of the degrees of freedom associated with a horizon as being ``deconfined''. Equation~\eqref{lutform} then indicates that the deficit in the Luttinger theorem is associated with deconfined degrees of freedom. This appears to resonate 
well with the many-body results ~\cite{FFL,Z2frac, Paramekanti2004} mentioned earlier, where the Luttinger deficit is indeed due to an emergent gauge symmetry. Note that to make the identification between a holographic system and these many-body states, one should identify the holographic gauge group with the emergent slave-particle gauge symmetry, in which case gauge-invariant fermionic operators can be interpreted as ``electrons'', and the deconfined horizon degrees of freedom can be interpreted as ``fractionalized'' excitations.   

 Defining 
\be \label{lude}
\sA \equiv \rho - \frac{1}{(2\pi)^{d-1}}\sum_i q_i V_i  
\ee 
 it is thus tempting to speculate that in a general system (not restricted to those with a gravity dual) the deficit $\sA$ provides a measure for the deconfined (or in the slave-particle context ``fractionalized'') degrees of freedom. For holographic systems $\sA$ then has an elegant geometric description~\eqref{adef1} in terms of electric
flux at the horizon.

\subsection{Deficit in the Magnus force}

We now consider the case where the $U(1)$ symmetry is broken by a charged condensate, i.e. we study a superfluid phase. Before turning to holographic theories, consider creating a vortex excitation in a {\it conventional} superfluid and moving it slowly through a closed loop of area $A$. In this process the vortex will pick up a Berry phase $\th$. In a conventional superfluid this phase is proportional to the total charge enclosed by the loop,  i.e. 
\be
\th = \frac{2\pi A}{q} \jT  \label{berry0}
\ee
This Berry phase actually means that there is a transverse force $F_T$ on a {\it moving} superfluid vortex. This is a well-known (if somewhat controversial) force, and in the literature is called the Magnus (and/or Iordanskii) force. In an ordinary superfluid one has
\be \label{mag0}
F_T^{i} = \frac{2\pi}{q}\ep^{ij}v_j \jT \ 
\ee 
 where $q$ is the charge of the condensate. 
 
However, it was found in~\cite{Paramekanti2004} that in certain exotic superfluid phases with an emergent $Z_2$  gauge symmetry, the Berry phase and the Magnus force exhibit a deficit compared with~\eqref{berry0} and~\eqref{mag0}, which has a similar origin to the Luttinger count deficit mentioned in the previous subsection. 

Keeping this in mind, we now turn to the holographic superfluid \cite{holographicsc}, in which the $U(1)$ symmetry is broken by the condensate of a charged scalar field in the bulk. In general there is also a horizon carrying a charge density. Interestingly, in such systems we also find that 
\be
\th = \frac{2\pi A}{q}\le(\jT - \sA\ri) \label{berry}
\ee
and 
\be
F_T^{i} = \frac{2\pi}{q}\ep^{ij}v_j\le(\jT - \sA\ri),
\ee
where $\sA$ is again given by~\eqref{adef1}.  That is,  the Magnus force also only picks up the charge outside the horizon. 

In parallel to the discussion of last subsection, defining 
\be \label{mnede}
\sA \equiv \rho - {q F_T \ov 2 \pi v}, 
\ee
it is again tempting to speculate that $\sA$ gives a measure of the deconfined charged degrees 
of freedom in a superfluid phase. 

It is rather intriguing that in the holographic context, the deficit~\eqref{lude} for the Luttinger theorem
and the deficit~\eqref{mnede} for the Magnus force of a superfluid vortex have precisely the same bulk origin in term of the electric flux~\eqref{adef1} at the horizon. As we will see in the main text, there is a common thread in the derivation of both deficits: it is the bulk Gauss's law, which relates the bulk electric field to {\it bulk} charge densities:
\be
\vec{\nabla} \cdot \vec{E}(r) = \rho_{bulk}(r) \ . \label{gauss}
\ee
It is the bulk electric field (evaluated at infinity) that determines the boundary theory charge density, but it is actually the bulk {\it charges} that contribute to the various observables that we compute. The horizon provides a mechanism for these two quantities to not agree, as will be more clear as we proceed. 

The plan of this paper is as follows: in Section \ref{sec:setup} we specify the sort of action that we will be working with and fix notation. In Section \ref{sec:fermi} we prove the modified Luttinger theorem \eqref{lutform} and apply it to various systems in the holographic literature. In Section \ref{sec:sf} we turn to the bosonic superfluid case. We provide a brief review of the physics of vortices in holographic superfluids and then derive the form of the Berry phase and Magnus force, again discussing how the results apply to various systems in the literature. Finally Section \ref{sec:conc} provides a brief discussion on the possible interpretations of these results. Various technical details are relegated to two appendices. 

\section{Setup} \label{sec:setup}
We will be holographically studying strongly-coupled $d$-dimensional field theories with a conserved $U(1)$ current $j^{\mu}$. We assume very little about this field theory, except that it has a limit in which it can be assumed to be dual to semiclassical gravity governed by the standard Einstein-Maxwell action
\be
S_{geom}[A,g] = \int d^{d+1} x \sqrt{-g}\le(\frac{1}{2\kappa^2} \le(R - 2\Lam\ri) -\frac{1}{4g_F^2} F^2\ri) \label{geom}
\ee
Here $F$ is the field strength tensor of a $U(1)$ gauge field $A$ that is dual to the conserved current $j^{\mu}$. $g_F$ is a bulk gauge coupling, which is assumed to be small. In each of the following sections we will add either fermonic or bosonic charged matter to this action. Note that we do not actually assume any particular form for the gravitational part of the action except that it support solutions with horizons and the usual sort of conformal boundary at infinity. The bulk gauge coupling $g_F^2$ can also be taken to depend on radius and so be the expectation value of a bulk scalar field, in which case our results apply to dilatonic systems, such as those studied in \cite{Goldstein:2009cv,arXiv:1111.2606}.

We assume that the bulk metric is diagonal, depending only on the holographic direction $r$ and preserving translational and rotational invariance. 
\be
ds^2 = -g_{tt} dt^2 + g_{rr} dr^2 + g_{ij} dx^i dx^j
\ee
Throughout this work $M,N$ will run over all the bulk directions and $\mu,\nu$ will run only over the field-theory spatial directions. While the fermionic portion of our discussion will apply to any dimension, in the scalar portion we will restrict do $d = 3$, so that the superfluid vortices are pointlike excitations. In the more interesting cases we will assume that the space-time has some sort of a horizon at $r = r_h$; this may be a degenerate or a finite-temperature horizon. 

In general, our notation for spinors is that of \cite{Faulkner:2009wj}. Gamma matrices with an underlined index (e.g. $\Ga^{\ur}$) are in an orthonormal frame. 

\section{Bulk fermions and Luttinger's theorem} \label{sec:fermi}
In this section we discuss the analog of Luttinger's theorem \eqref{lutform} as applied to holographic systems with charge carried by bulk fermions. Thus we add to the geometric part of the action \eqref{geom} bulk fermions,
\be
S_{\psi}[\psi,A] = -\int d^{d+1}x \sqrt{-g} i\bpsi (\Ga^M \sD_M - m) \psi \label{fermiac}
\ee
Here $\sD$ contains the spin connection as well as a coupling to the background gauge field with charge $q$. This bulk fermion $\psi(r,x)$ is dual to a charged fermionic operator which we will call $\sO(x)$. On many gravitational backgrounds \cite{Sachdev:2011ze,Liu:2009dm,Lee:2008xf,Cubrovic:2009ye,Faulkner:2009wj,Hartnoll:2011dm,Iqbal:2011in} the correlation functions of $\sO$ exhibit Fermi surfaces, i.e. near a shell in momentum space they takes the form
\be
G_R(\om,k) \sim \frac{Z}{\om - v_F(k-k_F) + \Sig(\om,k)} \ .
\ee
The precise form of the self-energy $\Sig(\om,k)$ depends on the infrared geometry of the gravitational background in question; see \cite{Iqbal:2011ae} for a review. We will show that for each $U(1)$ gauge symmetry in the bulk there exists a single constraint relating boundary theory charge densities with these Fermi surface singularities. A very similar result was recently obtained in \cite{Sachdev:2011ze} for a system with no horizon; our derivation is slightly more general in that we allow for a horizon and treat carefully the analytic structure and nontrivial quasiparticle widths associated with such a horizon. 
\subsection{Proof}
We would like to relate excitations of the bulk fermions to the bulk electric field. As we are dealing with fermions this will be an intrinsically quantum-mechanical contribution, and so we need to perform a one-loop calculation in the bulk. Note that the previous literature on the electron star \cite{Hartnoll:2010gu} should also be formally considered as a ``one-loop'' calculation, but in a fluid limit in which the scales determining the local electronic density of states were far higher than the curvature scale of the geometry; thus it is justified to treat each individual fermion as being arbitrarily well-localized, and the gas of fermions is treated as a fluid with the $d+1$-dimensional equation of state that is appropriate to flat space\footnote{This equation of state is the standard one from statistical-mechanics textbooks; of course if one wants to obtain it from the field theory of free fermions it does involve a one-loop fermion determinant.}. We will not make any such assumptions, and will treat the radial dependence of the fermion wavefunctions exactly. 

We will treat the gauge field and metric classically. We perform most of our calculation at zero temperature in Euclidean signature. The analytic continuation is 
\be
t = -i\tau \qquad \om = i\om_E \qquad iS = -S_E
\ee
where $S_E$ is the Euclidean action. We extend our results to finite temperature at the end. 

To begin, we simplify the spinor action by rescaling the bulk field as
\be
\psi = (-g g^{rr})^{-\frac{1}{4}} \Psi
\ee
As is well-documented \cite{Faulkner:2009wj}, this form eliminates the explicit appearance of the spin connection, and the Lorentzian spinor action in momentum space becomes 
\be
S_{\psi} = -\int dr \sqrt{g_{rr}} \frac{d\om}{2\pi}\frac{dk}{(2\pi)^{d-1}} \bPsi(\om,k;r) D \Psi(\om,k;r) \label{bspin}
\ee
where the bulk differential operator $D$ is
\be
D \equiv i\le(\Ga^{\ur}\p_r + \sqrt{g_{rr}}\le(-m + i \Ga^{\umu}K_{\mu}\sqrt{g^{\mu\mu}}\ri)\ri), \label{Ddef}
\ee
with $K_t = (-\om - A_t(r)),K_i = k_i$. This operator $D$ will have eigenfunctions:
\be
D\Psi_n(r,\om,k) = \lam_n \Psi_n(r,\om,k) \label{eigen}
\ee
Here the $\Psi_n$ are defined as usual by specifying boundary conditions on the spinor fields at both ends of the spacetime. The choice of boundary conditions on spinors at the AdS boundary is discussed in \cite{Iqbal:2009fd}. We will discuss the boundary conditions at the horizon as we need them. Note that eigenfunctions with $\lam_n = 0$ satisfy the bulk wave equation as well as the boundary conditions, and so are normalizable modes and thus signify a singularity in the {\it boundary}-theory Green's function (i.e. the correlation function of the dual operator $\sO$). 

\begin{figure}[h]
\begin{center}
\includegraphics[scale=0.50]{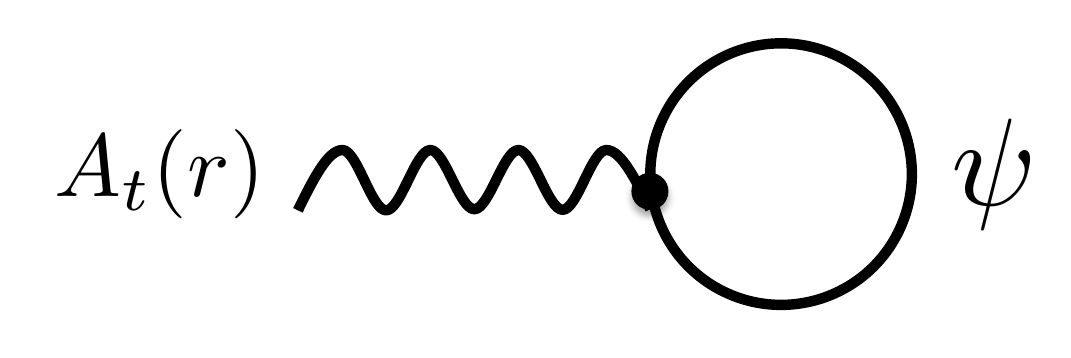}
\end{center}
\vskip -0.5cm
\caption{Feynman diagram representing one-loop fermion density.}
\label{fig:feynman}
\end{figure}

We now want to compute the effect of the fermion on the gauge field. Formally this is accomplished by integrating out the fermions, i.e. we work in Euclidean space and define a contribution to the effective action $\Ga[A]$ for the gauge field via
\be
\exp(-\Ga[A]) = \int [d\Psi] \exp(-S_E[\psi,A]) = (\det D), \label{effac}
\ee
where the determinant involves all modes in Euclidean signature, i.e. $D$ is evaluated with $\om$ purely on the imaginary axis. Thus the boundary conditions at the horizon for the eigenfunctions \eqref{eigen} are now fixed: we should demand that the wavefunctions be regular at the Euclidean horizon.

 We now want to compute the contribution of this quantum effective action to the classical equation of motion for $A_{\tau}$. Adding $\Ga[A]$ to the geometric part of the action \eqref{geom} and varying with respect to $A_{\tau}$ we find the following equation of motion:
\be
\frac{1}{g_F^2} \p_r(\sqrt{-g}F^{rt}) = i\frac{\delta \Ga[A]}{\delta A_{\tau}(r)} = -i\Tr D^{-1}\frac{\delta D}{\delta A_{\tau}(r)}\label{varac}
\ee
This is simply Gauss's law \eqref{gauss} in the bulk. The right-hand side is the pointwise charge density carried by the fermions, where in the last equality we have used \eqref{effac} to express it in terms of the operator $D$.  Note that this is the Feynman diagram shown in Figure \ref{fig:feynman}. We now need to evaluate this right-hand side. This can be done in several ways. In this main text we present a somewhat slick derivation using locality in the radial direction, and in Appendix \ref{spinorapp} we present a more pedestrian treatment using the spinor bulk-to-bulk propagator.

To begin, note that at fixed (Euclidean) frequency and momentum $D$ is an operator acting on functions $\Psi(r)$ that depend only on the radial direction. A basis for this function space is provided by the position eigenstates $|r\rangle$, where we can normalize to have the completeness relations
\be
\int_{r_h}^{\infty} dr \sqrt{g_{rr}} |r\rangle \langle r| = {\bf 1} \qquad \langle r | r' \rangle = \frac{1}{\sqrt{g_{rr}}}\delta(r - r')
\ee
Note that we assume that the basis is complete using only $r > r_h$, which is true provided that our function space is Euclidean frequency wavefunctions that are smooth at the horizon. Now $D$ is almost a local operator in $r$, and furthermore due to its gauge-invariance we can trade a derivative with respect to $A_{\tau}$ for a derivative with respect to $\om_E$ to find
\be
\le\langle r_1 \bigg| \frac{\delta D}{\delta A_{\tau}(r)}\bigg| r_2 \ri\rangle = -q\le\langle r_1 \bigg| \frac{\p D}{\p \om_E} \bigg| r_2 \ri\rangle \delta(r - r_2)
\ee
Inserting this representation into \eqref{varac} and performing some trivial manipulations we find the useful relation
\be
\frac{\delta \Ga[A]}{\delta A_{\tau}(r)} = -q\le\langle r \bigg| D^{-1}\frac{\p D}{\p \om_E} \bigg| r \ri\rangle \sqrt{g_{rr}}
\ee
This equation is valid at all points in the bulk but is somewhat formal and unenlightening; however if we insert this again into \eqref{varac} and integrate from the horizon to infinity then we find a trace over $D^{-1}\p_{\om_E}D$, which we can rewrite as a sum over the eigenvalue spectrum of $D$ \eqref{eigen}. The final expression is then
\begin{widetext}
\be
\frac{1}{g_F^2}\le(\sqrt{-g}F^{rt}\ri)\bigg|^{\infty}_{r_h} = -iq \int \frac{d\om_E}{(2\pi)}\frac{d^{d-1}k}{(2\pi)^{d-1}}  \sum_n \frac{\p}{\p\om_E}\log \lam_n(\om_E) \equiv I \ .\label{int1}
\ee
\end{widetext}
We have now removed the radial integral from the problem; the manipulations that follow are essentially those from regular field theory, although we will repackage them in a way that allows for easy generalization to multiple Fermi surfaces.

As we now explain, \eqref{int1} has something of a topological character. Note that the integration contour goes straight up the imaginary $\om$ axis. Imagine that we could close the contour on the left. Morally speaking, the integral is then counting the number of zeros of the eigenvalue spectrum in the left-half plane. Each of these zeros corresponds to a Fermi surface, and so the integral over $k$ will count the volume enclosed by the Fermi surfaces. This would have been a precise statement if the $\lam_n$ had been analytic functions of $\om$; however they actually contain singularities on the real $\om$ axis, and so to make this intuition precise we will need to work slightly harder.  

\begin{figure}[h]
\begin{center}
\includegraphics[scale=0.4]{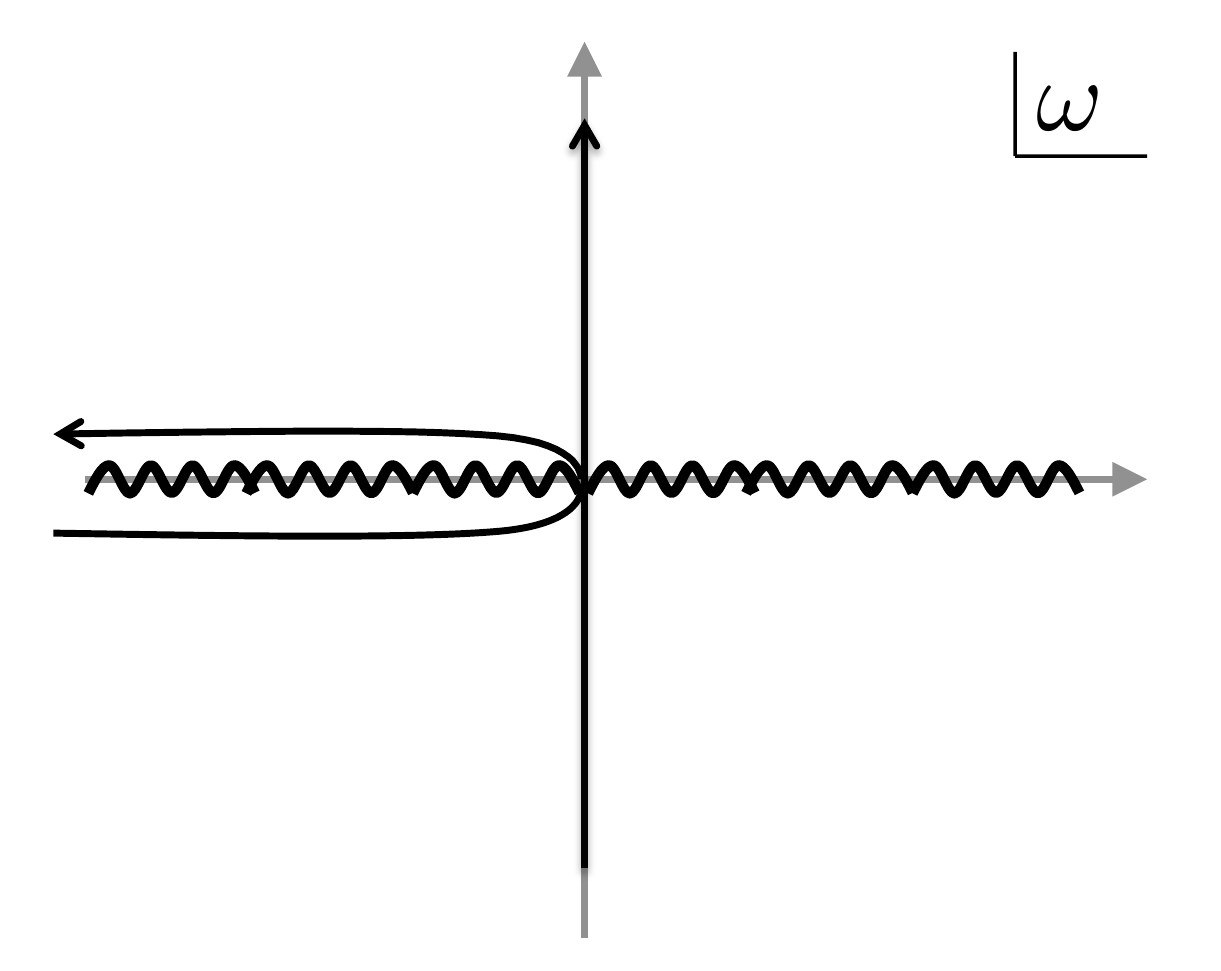}
\end{center}
\vskip -0.5cm
\caption{Contour manipulations used in evaluation of \eqref{int1}. Squiggles indicate non-analyticities in $\lam(\om)$ on real axis.}
\label{fig:contour1}
\end{figure}

Thus consider deforming the integration contour as shown in Figure \ref{fig:contour1} so that it wraps the negative real $\om$ axis. Now recall that $\lam(\om)$ is defined by analytic continuation from strictly imaginary $\om$; in the upper half plane the continuation to real $\om$ continues to {\it infalling} boundary conditions, defining a function $\lam^I(\om)$, whereas in the lower-half plane it continues to {\it outgoing} boundary conditions, defining a different complex function $\lam^O(\om)$ \cite{Iqbal:2009fd}. These conditions do not agree on the real $\om$ axis, leading to non-analyticities in $\lam(\om)$ there: thus we must use $\lam^I(\om)$ for the top half of the integral and $\lam^O(\om)$ for the bottom half. These functions are not unrelated; one can show that for fermions with time reversal symmetry $\lam^I(\om) = -\lam^O(\om)^*$. The integral \eqref{int1} can now be written as
\be
I = 2q\int\frac{d^{d-1}k}{(2\pi)^{d-1}} \int_{-\infty}^0\frac{d\om}{2\pi} \frac{\p}{\p\om}\al_k(\om),
\ee
where $\al_k(\om)$ is defined to be
\be
\al_k(\om) = \sum_n \arg \lam^I_n(\om) \ .
\ee
Now note that the integral essentially counts the winding in $\al_k(\om)$ from the origin to $-\infty$. As this is a topological expression, the full dependence of this expression on $k$ is expected to arise from the endpoint at $\om = 0$.\footnote{We assume the behavior at infinity is not affected by $k$.} Note further that $\lam_n(\om = 0)$ changes sign as $k$ is moved through each Fermi surface: thus as $k$ is increased $\al_k(0)$ jumps by $-\pi$ each time a Fermi surface is crossed. Thus we may write the integrand as
\be
I = q \int \frac{d^{d-1}k}{(2\pi)^{d-1}} N(k) \ .
\ee
where if we denote the set of Fermi surfaces by $\{k_F^{(i)}\}$, then $N(k)$ is a piecewise constant function:
\be
N(k) = \begin{cases} N_0 \qquad & k < k_F^{(1)} \\ N_0 - 1 \qquad & k_F^{(1)} < k < k_F^{(2)} \\ N_0 - 2 \qquad & k_F^{(2)} < k < k_F^{(3)} \\ \cdots, \\ 0 \qquad & k \to \infty  \end{cases} \label{nk},
\ee
where $N_0$ is a constant. In this expresion we {\it assume} that $N(k \to \infty) = 0$; the necessity for this will be shown shortly.
Denote by $V_{i}$ the volume enclosed by the $i$-th Fermi surface. Using the expression \eqref{nk}, we find
\begin{widetext}
\be
I = -\frac{q}{(2\pi)^{d-1}}\le(N_0 V_1 + (N_0 - 1)(V_2 - V_1) + (N_0 - 2)(V_3 - V_2) + \cdots\ri) = -\frac{q}{(2\pi)^{d-1}}\sum_i V_i \label{sharpI}
\ee
\end{widetext}
Somewhat magically, across each pair of terms the term involving $N_0$ vanishes, and the expression above reduces to the sum of Fermi surface volumes. Putting this back into \eqref{intanswer} and using the AdS/CFT expression \eqref{rhodef} for the boundary theory charge density, we find the desired result, i.e.  
\be
\frac{q}{(2\pi)^{d-1}}\sum_i V_i = \jT - \sA, \label{lutform2}
\ee
where $\jT$ is the total boundary theory charge density and $\sA$ is an anomalous term that measures the electric flux through the horizon,
\be
\sA = \le(-\frac{1}{g_F^2}\sqrt{-g}F^{rt}(r_h)\ri) \ . \label{adef}
\ee
As claimed above, the sum of Fermi surface volumes measures only the charge outside the horizon. 

Now note from \eqref{sharpI} that if $N(k \to \infty) \neq 0$, then the integral would receive a divergent contribution proportional to $N(\infty)$ multiplying the infinite volume in $k$-space that is outside the last Fermi surface. This seems like a clear UV pathology that should be avoided: by demanding that $N(\infty) = 0$ we are essentially asserting that there is no ``Fermi surface at infinity''. While this seems quite reasonable, this involves UV properties of the spectrum and we do not know how to prove this to be the case in general. Note from \eqref{nk} that this requires that $N_0$ be equal to the total number of Fermi surfaces. 

Before moving on, we outline a generalization; consider heating the system up to a finite temperature. If the field theory is in a deconfined phase we will have a black hole horizon in the interior; all excitations will acquire $T$-dependent widths, and all of the Fermi surfaces will become somewhat blurry. Nevertheless, there is still an exact Gauss's law at each point in the bulk, and so some analog of the above expression should still exist in the black hole geometry. 

It is shown in the Appendix that instead of \eqref{sharpI} one instead finds
\be
I(T) = -q\int \frac{d^{d-1}k}{(2\pi)^{d-1}} \sum_a \le(\ha - \frac{1}{\pi} \Im \psi\le(\ha + \frac{i\beta \Om_a(k)}{2\pi}\ri)\ri), \label{finT}
\ee
where $\psi$ is the digamma function and the sum over $a$ is now a sum over all of the complex quasinormal modes of the system $\Om_a(k)$. To understand the role of this digamma function, it is helpful to consider the case when the quasinormal modes are real, in which case one finds
\be
I(T) = -q \int \frac{d^{d-1}k}{(2\pi)^{d-1}} \sum_a \ha\le(1 - \tanh\le(\frac{\beta\Om_a(k)}{2}\ri)\ri), 
\ee
which is simply the familiar expression from Fermi liquid theory. Thus we see that the digamma function has the effect of smearing out the sharp expression seen in \eqref{sharpI}, and should be viewed as a generalization of the normal Fermi-Dirac distribution to complex frequencies. This is 
 in the spirit of \cite{Denef:2009kn,Denef:2009yy}, which showed that such generalizations play an important role in one-loop corrections to the thermodynamics of black holes.

\subsection{Applications}

We now discuss how the formula \eqref{lutform2} relates to various systems in the literature; most of the models discussed may be viewed as different solutions to the theory defined by the sum of \eqref{geom} and \eqref{fermiac}. Note that the derivation does not appear to be sensitive to the width of the excitation (provided that there is a normalizable state at $\om = 0$) and so this modified Luttinger's theorem should apply to non-Fermi liquid excitations as well as to stable quasiparticles. 

\subsubsection{Confining geometries}
We begin with the recent construction of \cite{Sachdev:2011ze}, in which the bulk spacetime is explicitly cutoff in the infrared to model a confining dual theory. In this case there is no horizon, and all the charge is clearly carried by fermions; $\sA = 0$ and the Luttinger count is saturated. Indeed the formula \eqref{lutform2} was previously explicitly derived in that work. It is argued there that the only gapless excitations are those associated with the Fermi surfaces, and this system is thus clearly a Fermi liquid.  

\subsubsection{Electron stars}
We now turn to the {\it electron star} geometry \cite{Hartnoll:2010gu}. In that solution there is no infrared cutoff; the space-time exists for arbitrarily large proper distance in the infrared, and is filled with a gas of fermions that sources the electric field. In a fluid limit for the fermions one can self-consistently solve for the radial dependence of the geometry, gauge field, and thermodynamic variables characterizing the fermions. This was done in \cite{Hartnoll:2009ns,Hartnoll:2010gu}, where it was found that in the infrared the geometry exhibits an emergent Lifshitz scaling, i.e. it takes the form
\be
ds^2 = L_l^2\le(\frac{-dt^2 + d\sig^2}{\sig^2} + \frac{d\vec{x}^2}{\sig^{\frac{2}{z}}}\ri) \qquad A = e_d \frac{dt}{\sig} \label{lifm}
\ee
This geometry is invariant under the scaling
\be
t \to \lam t \qquad x \to \lam^{\frac{1}{z}} x \qquad \sig \to \lam\sig,
\ee
where $z$ is an emergent dynamical critical exponent that can be related to the mass and charge of the bulk fermions. Note that the form of the gauge field is fixed by this scale invariance (up to the overall constant $e_d$). There is some sort of a degenerate horizon as we take $\sig \to \infty$; as we take $\sig \to 0$ eventually the geometry crosses over to an AdS$_{d+1}$ in the far UV.  

However, we can compute the electric field flux through this horizon; from \eqref{lifm} we find $\sA$ to be
\be
\sA \sim \frac{e_d}{g_F^2}\sig^{-\frac{d-1}{z}} \to 0 \label{avanish}
\ee
which thus vanishes as $\sig \to \infty$. Thus for all finite $z$ there is no flux carried by the degenerate horizon. By \eqref{lutform2} we see that the sum of all the boundary theory Fermi surfaces must make up the full charge density. This was verified explicitly in \cite{Hartnoll:2011dm,Iqbal:2011in}, who found by direct computation a dense set of boundary theory Fermi surfaces on the background whose infrared geometry is \eqref{lifm} and checked that indeed the sum of their volumes is equal to the total boundary theory charge density. Those calculations were performed in a WKB limit; now we see that the result must hold even away from this limit. 

Note that even though the geometry is very different in this case than the confining example above, from the point of view of the charged sector and the Luttinger count, there is little difference and it still appears that the charged degrees of freedom are ``confined''.\footnote{\HL{Note that  due to the warp factor $\sig^{-{2 \ov z}}$ before the spatial section of the metric~\eqref{lifm}, any bulk degrees of freedom with a nonzero boundary momentum will have a divergent local proper momentum approaching $\sig \to \infty$. Thus for gapless modes near a Fermi surface, the geometry~\eqref{lifm} essentially behaves as  a confining geometry with a ``soft'' wall.}} 

\subsubsection{Reissner-Nordstrom black hole}  \label{sec:RNBH}

In the previous examples the traditional Luttinger count was satisfied, as $\sA$ above was 0. We now turn to the opposite extreme, which is one of the workhorses of finite-density holography: the well-studied charged Reissner-Nordstrom-AdS black hole \cite{RNBH}. This gravity background is a solution to the pure Einstein-Maxwell part of the action \eqref{geom}, with no fermions sourcing the background. The infrared geometry can be understood as the $z \to \infty$ limit of \eqref{lifm}, i.e. AdS$_2 \times \mathbb{R}^{d-1}$: 
\be
ds^2 = L_2^2\le(\frac{-dt^2 + d\sig^2}{\sig^2}\ri) + a^2 d\vec{x}^2 \qquad A = e_d \frac{dt}{\sig} 
\ee
This horizon carries a charge that appears independent of the fermion field content, i.e. we have
\be
\sA = \frac{e_d}{g_F^2}\frac{a^{d-1}}{L_2^2}
\ee
At the level of classical gravity, this is the {\it full} contribution to the boundary theory charge density; if we were to obtain this bulk action from an actual stringy embedding, it would scale as $O(N^2)$. The fermions play no role in this analysis; thus in the extreme large $N$ limit, this is the opposite limit to the one considered above; all of the charge is behind the horizon, and the traditional Luttinger count is thus maximally violated. 

Nevertheless, if one probes this charge density with a bulk fermionic field \eqref{fermiac}, one finds that the correlation functions of the boundary theory fermion operator contains a small (i.e. $O(1)$) number of isolated Fermi surfaces \cite{Liu:2009dm,Lee:2008xf,Cubrovic:2009ye,Faulkner:2009wj}. These take the form
\be
G_R(\om,k) \sim \frac{Z}{\om - v_F(k - k_F) + c \om^{2\nu}} \label{nfl}
\ee
with $\nu$ a nontrivial scaling exponent \cite{Faulkner:2009wj} that is related to the scaling dimension of the fermion operator in the AdS$_2$ region. One expects that these modes should also contribute to the total charge density. This section may then be viewed as a calculation of this small correction, which from \eqref{lutform2} can be seen to be of $O(1)$ rather than $O(\frac{1}{g_F^2}) \sim O(N^2)$.  

Over most of the parameter space the system also often contains a set of incoherent excitations dubbed the ``oscillatory region'' by \cite{Liu:2009dm,Faulkner:2009wj}. Their existence can be traced to the AdS$_2$ scaling dimension going {\it imaginary}. This set of excitations is quite peculiar; for example at zero temperature and frequency one finds that for all $k$ less than a threshold value $k_0$:
\be
\Im G_R(\om = 0; k < k_0) \neq 0,
\ee
i.e. even at precisely zero frequency there is nonvanishing spectral weight. Related to this, for $k < k_0$ one finds an {\it infinite} number of poles in the retarded Green's function, geometrically spaced along a line in the lower-half plane:
\be
\Om^*_n(k) = \Om_0 \exp\le(i\th - \frac{n\pi}{\lam_k}\ri) \qquad n \in \mathbb{Z^+},
\ee
where $\Om_0$ is a UV scale related to the chemical potential, $\Om_0 \sim \mu$ and $\lam_k$ is related to the imaginary part of the IR conformal dimension of the fermion operator.  

One should now include the contribution of these excitations to the one-loop calculation done above. It is not at all clear how to do this in a controlled manner, but to obtain a crude estimate we can imagine going to finite but very small temperature. Then the spectrum is cut off at frequencies $\om \sim T$, and one finds that the total number of poles is $N \sim \log\le(\frac{\mu}{T}\ri)$. Each of these poles will contribute to the answer via the formula \eqref{finT}; it is difficult to estimate the contribution of each pole, but assuming it to be a finite number $\#$ we find for the contribution to the oscillatory region
\be
I_{osc}(T) \sim \#\log\le(\frac{\mu}{T}\ri)
\ee
Note the $T \to 0$ divergence. This should be added to the ``anomalous'' contribution from classical gravity $\sA$ to obtain the full field theory charge density as in \eqref{lutform2}. It is clear that at exponentially low temperatures this quantum correction will outweigh the classical horizon contribution and this calculation will break down: what the system is telling us is that we need to take into account the effect of the fermions on the classical electric field (and thus also on the geometry). 

This conclusion was first reached in \cite{Hartnoll:2009ns} via an argument that is essentially a WKB limit of the calculation performed above. Once we include this backreaction effect, the final geometry is that of the electron star~\eqref{lifm}. In particular, this set of incoherent excitations is resolved into a dense (but discrete) set of Fermi surfaces -- these excitations suck the charge out of the horizon and saturate the Luttinger count. \HL{As emphasized in~\cite{Iqbal:2011in}, this is an example in which going to low energies one finds a smooth crossover from a ``deconfined'' phase in which charges are described by a horizon to a ``confined'' phase where charges live outside the horizon.}\footnote{\HL{Here by ``confined'' and ``deconfined'' we only refer to charge degrees of freedom. } }

\subsubsection{Dilatonic systems}
We now turn to systems that have a nontrivial kinetic term for the bulk gauge field, i.e. where the gauge coupling is essentially taken to  be a function of a dilaton $\phi$ that can run:
\be
S[A] = -\frac{1}{4}\int d^{d+1}x \sqrt{-g} Z(\phi) F^2 \ . \label{dilatonic}
\ee
Generally $\phi(r)$ will have a nontrivial profile in the infrared, allowing for more interesting infrared geometries than those discussed above. Note first of all that our derivation above never assumed that the gauge coupling $g_F^2$ was constant in $r$, and so all of the discussion can be immediately taken over, except that the electric field flux now depends explicitly on $\phi$, e.g. instead of \eqref{adef} we have
\be
\sA = \le(\sqrt{-g}Z(\phi)F^{rt}\ri)\bigg|^{\infty}_{r_h}
\ee
and so on in all relevant formulas. 

We discuss first the dilatonic black holes studied in \cite{Goldstein:2009cv}. These are solutions to the action \eqref{dilatonic} together with the Einstein-Hilbert term and a kinetic term for $\phi$, but no fermions. They have an infrared geometry whose metric is Lifshitz with finite $z$ as in \eqref{lifm}, but the dilaton $\phi$ runs away in the infrared in such a way that the flux through the horizon $\sA$ is finite in the infrared, rather than the vanishing result found in \eqref{avanish}. Thus despite the very different infrared geometry from the Reissner-Nordstrom case, from the point of view of the anomalous Luttinger count the systems appear very similar; they also maximally violate the Luttinger count.

Finally, we turn to the construction of ``fractionalization of holographic Fermi surfaces'' in \cite{arXiv:1111.2606}, who study a gauge field action of the form \eqref{dilatonic} together with fermions. In this construction one finds that generically there is a nonzero charge both behind the horizon (i.e. in $\sA \neq 0$) {\it and} in a fermion density outside. The fraction of charge behind the horizon can be tuned as a function of boundary theory couplings, allowing one to interpolate between a phase that saturates the traditional Luttinger count and one that maximally violates it. This is the first zero temperature example in which none of the terms in \eqref{lutform2} is zero, and so is a nontrivial application of this formula. 
\section{Vortices in holographic superfluids}\label{sec:sf}
In the previous section we studied fermions that carry charge in the bulk, and we explained how there exists a field-theoretical observable (the sum of the Fermi surface volumes) that measures only this bulk charge, and is insensitive to the electric field flux behind the horizon. In this section we seek a similar observable for the case when the bulk charge is carried by {\it bosons}. In this case generically the bulk boson condenses, breaking the $U(1)$ symmetry, and we find ourselves in a superfluid state \cite{holographicsc}. See \cite{Hartnoll:2009sz,Horowitz:2010gk,Herzog:2009xv} for reviews discussing the physics of such a phase. It is perhaps not immediately clear what the relevant observable should be in this case. 

A clue is given by \cite{Paramekanti2004}, who study phases of matter with an emergent $Z_2$ gauge symmetry. In the condensed matter context, such emergent gauge structures in {\it fermionic} systems are often associated with violations of Luttinger's theorem \cite{FFL,Z2frac}. There also exist superfluid states with such an order: one of the results of \cite{Paramekanti2004} is that {\it vortex} excitations in such systems provide an interesting probe of the charge density. 

To understand this, consider first a generic (conventional) many-body system at zero temperature in two spatial dimensions, in which the ground state is a superfluid that breaks a $U(1)$ symmetry. Such a state will generically have gapped vortex excitations, around which the fluid circulation is quantized in multiples of $\frac{2\pi}{q}$, where $q$ is the charge of the boson that is condensed. 

Now consider moving this vortex in a closed loop enclosing an area $A$. It is a standard result \cite{Haldane1985,Ao1993} that the wavefunction of the system will pick up a Berry phase $\th$ that is proportional to the area enclosed, and is
\be
\th = \frac{2\pi}{q} \rho A \label{normB}
\ee
where $\rho$ is the total number density, which at zero temperature is stored entirely in the condensate. In other words, in a conventional system the Berry phase accumulated by the vortex counts the total charge density enclosed by the loop.   

There is a macroscopic manifestation of this Berry phase. To understand this, it is helpful to consider the familiar case of a charged particle moving in a magnetic field $\vec{B} = \vec{\nabla} \times \vec{a}$. If this particle is moved through a closed loop $\Ga$, it accumulates a Berry phase that is
\be
\th_{B} = q\oint_{\Ga} d\vec{x} \cdot \vec{a} =q B A \label{magfield}
\ee
i.e. the phase counts the flux enclosed by the loop. Indeed, reversing the logic of \eqref{magfield}, the presence of such a phase implies a term in the classical Lagrangian that is linear in the velocity, which means that a particle moving in a magnetic field with velocity $v$ feels a Lorentz force $F_L = q \vec{v} \times \vec{B}$. By analogy, we conclude that the result \eqref{normB} suggests that a vortex moving with velocity $v$ feels a transverse force proportional to the density,
\be
F_T^i = \frac{2\pi}{q} \rho \ep^{ij} v_j \ .
\ee
This is called the Magnus force. This transverse force on a region of circulating fluid exists in any hydrodynamic system and is not restricted to superfluids, and in everyday life is responsible for such important effects as the motion of a curve ball and the lift force on an airplane wing.

Now consider heating the system up to a finite temperature. The Berry phase ceases to be well-defined; nevertheless one expects the transverse force on a vortex to remain a well-defined observable. The precise form of this force is now a matter of considerable controversy \cite{controversy}. Basically, at nonzero temperatures the density has both a normal component $\rho_n$ and a superfluid component $\rho_s$. This situation has been well-studied in the context of Galilean superfluids, where the normal component comes from thermally excited phonons. It is not clear whether or not (and how much of) this normal component contributes: see Chapter 2 of \cite{lara} for a recent review. We do not attempt to review the debate here, but we point out that for that if the fluid is at rest, then there exists a set of arguments that lead one to conclude
\be
F_T(T)^i = \frac{2\pi}{q} \le(\rho_s + \rho_n\ri) \ep^{ij} v_j = \frac{2\pi}{q} \rho \ep^{ij} v_j \label{cont}
\ee
i.e. both the normal and superfluid components contribute\footnote{The part proportional to $\rho_s$ is called the Magnus force; the part proportional to $\rho_n$ is conventionally called the Iordanskii force.}. There exist other arguments that lead to other results: we mention this result here because we believe that our holographic calculation can be interpreted as lending some support to it, although in a somewhat subtle way. 

For relativistic superfluids without an underlying particle description the situation appears more confusing still, as the separation of the total charge density into a normal and superfluid component is no longer obvious. While there does exist a prescription for performing such a separation (see e.g. \cite{hep-ph/0011246,arXiv:0809.4870,arXiv:0906.4810} for a recent discussion or a brief review in Appendix \ref{scalarapp}), it does not appear that the holographic calculation of the Magnus force is sensitive to the difference between the two, as will be more clear as we proceed. \footnote{Note also that in the Galilean superfluid the symmetry that is broken -- number density -- is actually entangled with spacetime symmetry generators in a nontrivial way: e.g. the momentum is proportional to the number current, which is not the case in the relativistic system under study here. We are thus not entirely certain how easily results from the Galilean superfluid should be taken over to the relativistic setting, and one should perhaps use caution when comparing the two. Unfortunately, we are not aware of a careful relativistic field-theoretical Magnus force calculation to which we can compare our holographic results. We thank K.~Balasubramanian for discussions in this regard.}

In any case, these results do suggest that superfluid vortices form a probe of the charge density. In the remainder of this section we will compute both the Berry phase and the transverse force felt by a vortex in a holographic superfluid with a horizon. Just as in the fermionic example above, we will find that only the charge {\it outside} the horizon contributes to these observables. At the end we will comment on how our results relate to the controversy. 

\subsection{Background on holographic vortices}
In this section we will be working with a $2+1$-dimensional field theory so that vortices are point like excitations. We will work with a simple bulk action, in which we add to \eqref{geom} a bulk scalar field with charge $q$,
\be
S_\phi[\phi,A] = -\int d^{4}x \sqrt{-g}\le( (D\phi)^{\dagger} D\phi + V(\phi^{\dagger}\phi)\ri) \label{scalarac}
\ee
This is the action of the well-studied holographic superfluid \cite{holographicsc}, and it is well-known that at sufficiently low temperatures (compared to the chemical potential) the system generally orders into a superfluid state where the scalar $\phi$ obtains a nontrivial bulk profile:
\be
\phi(r) = v_0(r), \label{bgs}
\ee
breaking the $U(1)$ symmetry in the bulk. 

$A_t(r)$ also has a nontrivial profile, indicating (via \eqref{rhodef}) that the field theory state is at nonzero charge density. Some fraction of this charge is stored in the condensate outside the horizon. At finite temperatures, there is also a nonzero electric field flux through the horizon. These both contribute to the total charge density.

The detailed thermodynamics at very low temperatures depends on the bulk potential, but we will remain at finite temperature and assume that the potential is such that this condensation occurs. All of our calculations will be done in a probe limit where the scalar and gauge field do not backreact on the geometry; however we do expect the essential features of our discussion to survive the inclusion of back reaction. One may assume the background geometry to be just the AdS-Schwarzschild black brane; we have a horizon at radius $r = r_h$. 

Now on general grounds whenever we break such a $U(1)$ gauge symmetry we expect to find a state corresponding to an Abrikosov string where the phase of the condensate winds through $2\pi$ around a line in the bulk. In holographic models such solutions have been explicitly constructed \cite{Albash:2009iq,Montull:2009fe, Keranen:2009re, Maeda:2009vf} by numerically solving the relevant partial differential equations. Many features of these solutions may be understood analytically: imagine such a vortex line localized in the field theory directions at a location $\vec{x}_0$, stretching from the AdS boundary down to the horizon at $r = r_h$. Let us use polar coordinates $(\rho,\th)$ for the field-theory spatial directions, centering the origin at $\vec{x} = \vec{x_0}$. We expect the bulk fields to take the form
\be
\phi(x) = v(r,\rho)e^{i\al(\th)} \qquad A_{\th}(\rho \to \infty) \sim \frac{1}{q} + O\le(\frac{1}{\rho}\ri) \label{soliton}
\ee
The profile $v(r,\rho)$ will be fixed by bulk dynamics; it must vanish at the vortex core $\rho = 0$ and far from the vortex it should approach the background profile \eqref{bgs}. As usual, the condition on $A_{\th}$ is found by appealing to finiteness of the energy far from the vortex by demanding that the integral of the term $g^{\th\th}(\p_{\th}\al - qA_{\th})^2$ in the energy of the vortex be finite (see e.g. \cite{coleman}). This means that the flux enclosed in the string is
\be
\Phi = \oint d\vec{s} \cdot \vec{A} = \frac{2\pi}{q} \ . \label{flux}
\ee
Note that outside the vortex core nothing is sourcing $A_{\th}$, so it is essentially ``pure gauge'', where the quotes indicate that the gauge parameter of which it is a derivative is not a single valued function. We can rewrite it as
\be
A_M(x) = \frac{1}{q}\p_M \Lam(x;x_0) \qquad \Lam(x;x_0) = \arg(\vec{x} - \vec{x}_0) \label{vortex}
\ee
Thus as we move in a circle around the vortex $\Lam$ sweeps through $2\pi$. 

It is clear that the field-theory interpretation of this configuration is a pointlike vortex excitation located at $\vec{x}_0$. There are some subtleties here associated with boundary conditions, which we review in Appendix \ref{scalarapp}. 

\subsection{Berry phase}
\begin{figure}[h]
\begin{center}
\includegraphics[scale=0.60]{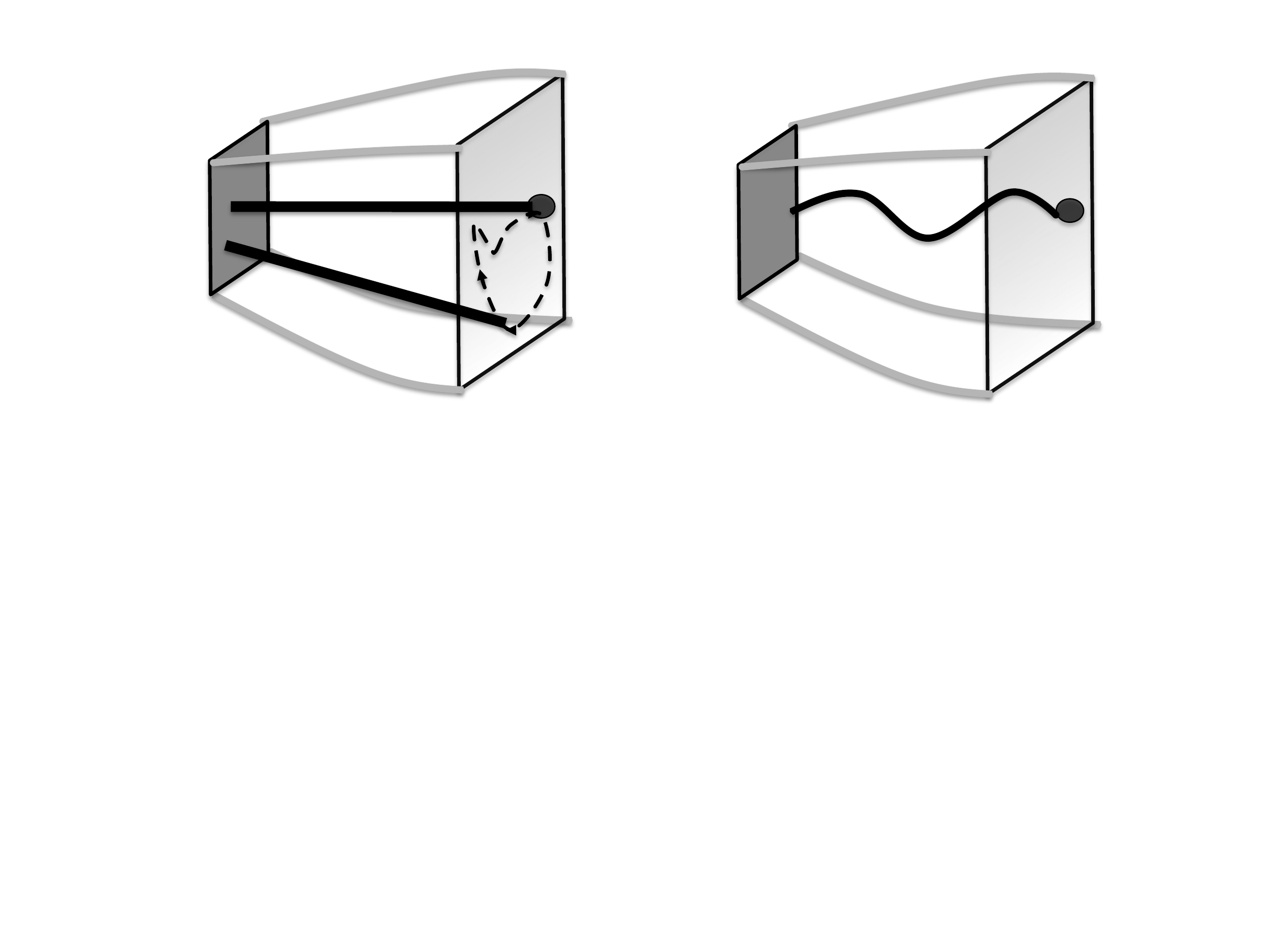}
\end{center}
\vskip -0.5cm
\caption{Moving vortex slowly in a closed loop. The Berry phase counts the bulk charge enclosed by the loop in the bulk.}
\label{fig:vortexarea}
\end{figure}

We now attempt a naive gravity computation of the Berry phase accumulated by this vortex if we move it very slowly in a closed loop, i.e. make $\vec{x}_0(t)$ a slow function of $t$, as in Figure \ref{fig:vortexarea}. The final answer is simple to understand; the vortex carries magnetic flux $\frac{2\pi}{q}$. If we move it through a closed loop in the bulk, all the bulk charge density that is enclosed in the loop will encircle the flux tube and contribute a phase via the usual Aharonov-Bohm effect. Thus we naively expect the Berry phase to only notice the charge outside the horizon. 

We will compute this phase by evaluating the bulk Lorentzian action on the adiabatic trajectory. First, we note that the key term in the action is  
\be
S_{int} = \int d^4x \sqrt{-g} A_{M}\sJ^{M}
\ee
Here $\sJ^{M}$ is the {\it bulk} current, built from the usual Noether procedure applied to \eqref{scalarac}. Now there is a background value of $A_t^{(0)}$ that is sourced by the superfluid profile; we will neglect this term as it does not change as we move the vortex, and from now on one should assume that $A_t$ is {\it purely the vortex contribution} \eqref{vortex}. On the other hand the background value of $\sJ^t$ will be important. In the limit of adiabatic transport the value of $\sJ^i$ is $0$, and so from \eqref{vortex} we have
\be
S_{int} = \frac{1}{q}\int d^3x  \sqrt{-g} \int dt \frac{d}{dt} \Lam(x;x_0(t)) \sJ^t 
\ee
Now consider this expression carefully. The integral over $t$ simply measures the winding of $\Lam$ as we move the vortex in a closed loop. If $\vec{x}$ is {\it inside} the closed loop $\vec{x}_0(t)$, then $\Lam$ winds through $2\pi$. If it is {\it outside} the closed loop then the winding is $0$. Thus the integral receives contributions only from points inside the loop, and is
\be
S_{int} = \frac{2\pi A}{q} \int dr \sqrt{-g} \sJ^t(r) \label{sint}
\ee 
where $A$ is the transverse area enclosed by the loop, i.e. the phase is proportional to the bulk charge enclosed. 

We now relate this to a boundary theory quantity. Consider the bulk Gauss's law:
\be
\p_r\le(\frac{\sqrt{-g}}{g_F^2}F^{rt}\ri) = -\sqrt{-g}\sJ^t \label{gaussbulk}
\ee
Again, integrate this equation from the horizon to the boundary and use \eqref{rhodef}; we find then that the Berry phase is
\be
\th = S_{int} = \frac{2\pi A}{q}\le(\jT - \sA\ri), \label{sfluttv}
\ee
where as before $\sA$ is the electric flux through the horizon \eqref{adef}. This is the claimed result. Note that this is precisely the same pattern that we find in the Luttinger count violation in the fermionic case: the bulk observable is sensitive only to the charge {\it outside} the horizon, and so $\sA$ should be viewed as an anomalous contribution.

Let us now turn a critical eye to this computation. From the bulk point of view, we assumed that the string was moving {\it rigidly} as we moved it from the boundary. However in a spacetime with a horizon, this is clearly impossible. No matter how slowly we may move the string at infinity, from the point of view of an observer at infinity, the movement will never propagate through to the horizon; the pulses of movement will slow down and appear to freeze just outside. There is some confusion as to how we should deal with the interaction of the string with the horizon. This confusion is dual to the field-theoretical statement that in a system with a horizon and the associated gapless degrees of freedom, there is no such thing as ``adiabatic'' transport, and so the idea of the Berry phase itself is not very well-defined. 

Thus while the calculation illustrated above is entertaining, it should not be taken too seriously. 

\begin{figure}[h]
\begin{center}
\includegraphics[scale=0.60]{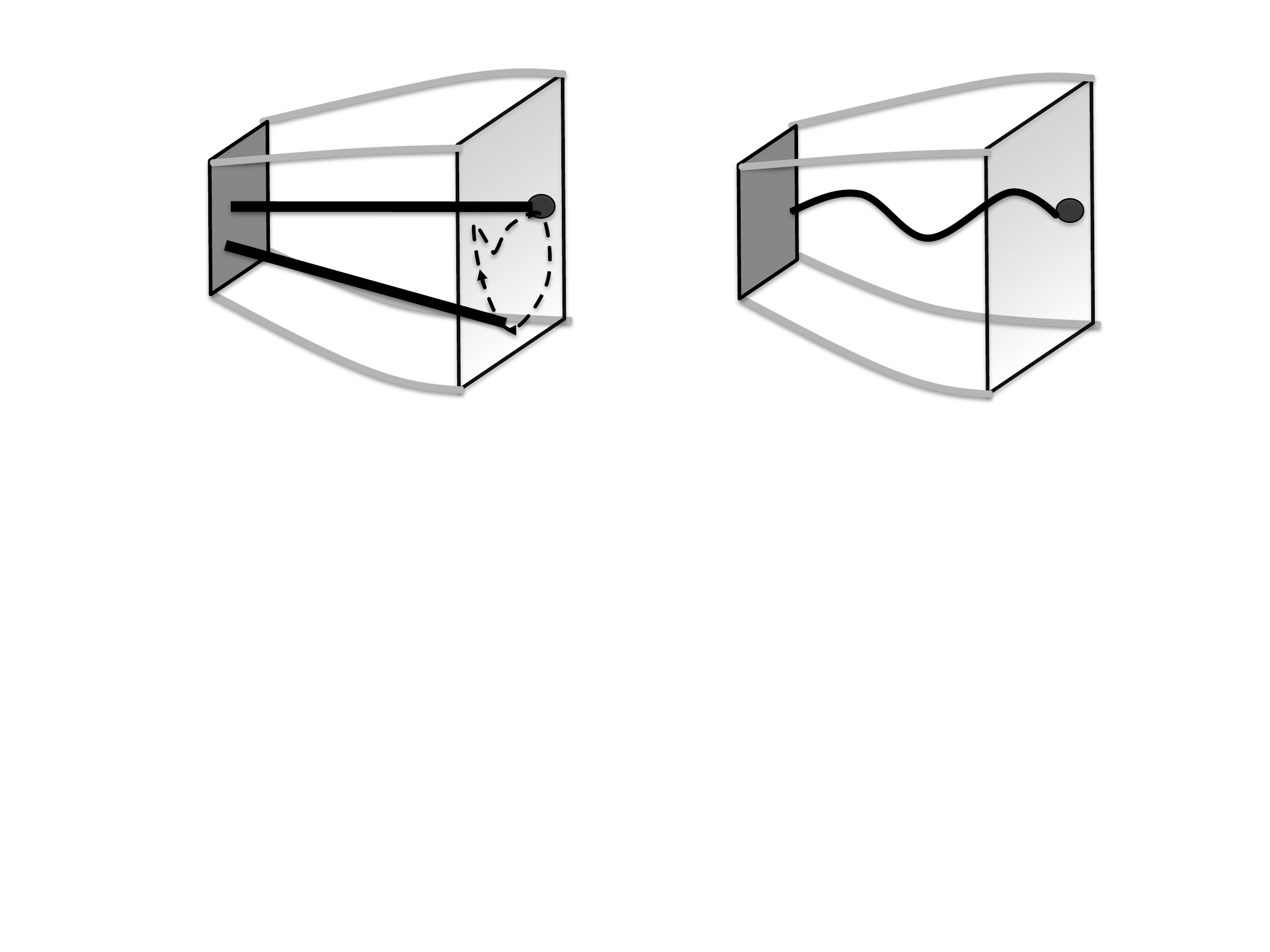}
\end{center}
\vskip -0.5cm
\caption{Sending ripples down vortex string to compute Magnus force in linear response.}
\label{fig:wiggly}
\end{figure}

\subsection{Magnus force}
We will now compute an observable that should be well-defined even at finite temperature; the transverse force on a vortex moving with a velocity $\vec{v}$. We will do this by sending low frequency ripples down the vortex string into the horizon as in Figure \ref{fig:wiggly} and computing the force that must be supplied to sustain such a ripple. We shall see that this can be formulated as a linear response problem. To do this we need to describe {\it dynamics} along the vortex, i.e. we seek an action that describes ripples in the collective coordinates characterizing the solution \eqref{soliton}. We assume that such an action is given by
\be
S_{vortex} = S_{NG} + S_{int}
\ee
where $S_{NG}$ is the Nambu-Goto action
\be
S = -\int d^{2}\sig\;T_p(X)\sqrt{\det\le(\p_{a}X^{M}\p_{B} X^N\ri)g_{MN}(X)}
\ee
Here we treat the core of the vortex as a stringlike excitation with a tension $T_p$ that can vary along the radial (holographic) direction. We pick coordinates on the world sheet $(r,t)$ that correspond to the embedding coordinates in spacetime; then the quadratic term for fluctuations in the transverse $X^i$ is
\begin{widetext}
\be
S_{quad} = -\frac{1}{2} \int dr dt T_p(r)\sqrt{g_{rr}g_{tt}}g_{xx} \le(g^{rr}(\p_r X)^2 + g^{tt}(\p_t X)^2\ri)
\ee
\end{widetext}

Let us now consider the boundary conditions on the string. Consider the canonical momentum of the string coordinate with respect to a foliation in the $r$-direction, i.e. at the quadratic level we have
\be
\Pi_i \equiv -T_p\le(\sqrt{-g}g^{rr}\p_r X_{i}\ri) \ .
\ee
Note that if we are applying a force $F_i$ on the string at the UV boundary, the value of this canonical momentum at infinity should be equal to the force applied, i.e.
\be
\Pi_i(r \to \infty) = F_i \ . \label{fdef}
\ee
When $F_i = 0$ this is the normal Neumann boundary condition. This forced result is familiar in the context of the classic string dragging calculations in AdS/CFT \cite{Herzog:2006gh,Gubser:2006bz}. Thus to compute the force on the vortex we should compute this canonical momentum at infinity. 

We turn now to the other term in the action. As before, $S_{int}$ describes the interaction of the vortex with the background charge density and is given by \eqref{sint}, except that now $A_M$ depends explicitly on the location of the string, i.e.
\be
A_{M} = \frac{1}{q}\frac{\p}{\p x^M} \Lam(x;X) \qquad \Lam(x;X) = \arg\le(\vec{x} - \vec{X}\ri)
\ee
Note now that as only $\sJ^t$ is nonzero, the action is only sensitive to $A_t$, which we write via the chain rule as
\be
A_t = \frac{1}{q}\dot{X}^i\frac{\p}{\p X^i}\Lam(x;X) 
\ee
Thus the term in the action may be written as
\begin{widetext}
\be
S_{int} = \int dr dt \sqrt{g_{rr} g_{tt}}\dot{X}^i a_i \qquad a_i \equiv \frac{1}{q}\int d^2 x g_{xx} \frac{\p}{\p X^i}\Lam(x;X) \sJ^t\ . 
\label{effaction}
\ee
Here we have defined an effective vector potential $a_i$, which couples to each bit of the string the same way as a vector potential couples to a charged particle.  Note that this action appears nonlocal, as it appears to involve an integral over the full field-theory spatial directions, and not just an integral over the string worldsheet\footnote{Note also that we are also neglecting any effects from the core of the vortex; one should assume that the fluctuations $X^i$ are larger than the core but still small enough to be characterized by linear response.}.  We will show that the equations of motion are however local.

Now we vary $S_{int}$. We find after some algebra
\be
\delta S_{int} = \int dr dt \sqrt{g_{rr}g_{tt}}\sJ^t \le[(\delta X \dot{Y} - \delta Y \dot{X})\le(\frac{\p a_y}{\p X} - \frac{\p a_x}{\p Y}  \ri) \ri]\ee
This equation is valid for any vector potential $a_i$ (note that it is simply taking the curl). In our case, $a_i$ appears to be the derivative of a scalar function $\Lam(x;X)$ and naively the curl would vanish; but of course the scalar function is not single-valued, and now we use the following identity:
\be
[\p_X, \p_Y]\Lam(x;X) = (2\pi) \delta^{(2)}\le(\vec{x} - \vec{X}\ri)
\ee
to convert the integrand into a delta function in the spatial directions. Thus the equations of motion are local on the string worldsheet. The final equations of motion are
\begin{align}
\p_r\le(T_p \sqrt{-g}g^{rr}\p_r X^i\ri) + T_p \sqrt{-g}g^{tt}\p_t^2 X^i + \ep^{ij}\frac{2\pi}{q}\sqrt{-g} \sJ^t \p_t X^j & =  0 \ .
\end{align}
\end{widetext}
This is the massless scalar wave equation plus an extra mixing term. 

We may decouple these equations by introducing the linear combinations and their canonical momenta
\be
X_{\pm} \equiv X \pm i Y \qquad \Pi_{\pm} \equiv -T_p\le(\sqrt{-g}g^{rr}\p_r X_{\pm}\ri),
\ee
after which we find
\be
\p_{r}\Pi_{\pm} + \sqrt{-g}g^{tt}\om^2 T_p X_{\pm} \pm \om \frac{2\pi}{q}\sqrt{-g} \sJ^t X_{\pm} = 0 \label{eq1}
\ee 
The above manipulations were valid on any background. We now specialize to a finite-temperature horizon. An efficient method for the solution of equations such as \eqref{eq1} on such backgrounds was described in \cite{Iqbal:2008by}. We do not review the method here; we just point that out a convenient object to consider is the following ratio:
\be
\sig_{\pm} \equiv \frac{\Pi_{\pm}}{i\om X_\pm} \label{sigdef}
\ee
It is shown in \cite{Iqbal:2008by} that horizon boundary conditions force this to be a constant (call it $\sig(r_h)$) at the horizon. In our context, when evaluated at infinity it will relate the applied force to the displacement of the vortex via \eqref{fdef}\footnote{If $X^i$ were a bulk gauge or gravitational mode, this same object $\sigma(r \to \infty)$ would have been a transport coefficient \cite{Iqbal:2008by}. Indeed there is a great deal of formal similarity between the Magnus force computation described here and a holographic computation of a Hall conductivity; it would be interesting to understand if this is related to some form of particle-vortex duality in the boundary theory.}. 
From \eqref{eq1} we find the radial evolution equation:
\be
\p_r \sig_{\pm} = \pm \frac{2\pi i}{q} \sqrt{-g} \sJ^t + i\om \le(T_p \sqrt{-g}g^{tt} + \frac{\sig_{\pm}^2}{T_p \sqrt{-g} g^{rr}}\ri)
\ee
and now finally we go to the low-frequency limit, which lets us discard the second term. We can then immediately integrate to find the following answer
\be
\sig_{\pm}(\infty) = \sig(r_h) \pm \frac{2\pi i}{q} \int_{r_h}^{\infty} dr \sqrt{-g}\sJ^t \equiv \sig(r_h) \pm iM \ .
\ee
$M$ counts the charge outside the horizon; as usual, we use the bulk Gauss's law \eqref{gaussbulk} to express it in terms of field theory quantities as 
\be
M = \le(\frac{2\pi}{q}\int_{r_h}^{\infty} dr \sqrt{-g}\sJ^t\ri) = \frac{2\pi}{q}(\rho - \sA) \ .
\ee
The interpretation of this equation is far more clear if we convert back to the $(X,Y)$ basis and use \eqref{fdef} to relate the canonical momentum at infinity to the applied force. We then find
\be
F^i_T = \ep^{ij} \frac{2\pi}{q}(\rho - \sA)\frac{d}{dt}X^j \label{magnus}
\ee
We see that this off-diagonal component of the force is proportional to the velocity, and is precisely the Magnus force. There is also a diagonal component proportional to $\sig(r_h)$ that is purely dissipative and that we have not written down. \eqref{magnus} is the desired result. 

\subsection{Applications}
We now discuss how these results apply to various systems in the literature. 

\subsubsection{Finite temperature superfluids}
We first study the usual holographic superfluid, i.e. the model studied in \cite{holographicsc}. For reasons to be clear below we call this the {\it deconfined} holographic superfluid. All of the results above apply directly to this model; we see that the coefficient of the Magnus force \eqref{magnus} in fact agrees with the naive Berry phase calculation \eqref{sfluttv}. The coefficient of this force thus directly measures the charge outside the horizon. At finite temperatures this will disagree with the total charge density by the anomalous term $\sA$; we point out that the structure of the violation is identical to that seen in the rather different Luttinger count calculation for fermionic systems. 

We now relate this to the controversy discussed earlier. As mentioned earlier, in conventional Galilean superfluids it is a somewhat controversial matter whether or not the {\it normal} component of the number density contributes to the transverse force \cite{controversy}. The normal density in question is a thermally excited phonon gas, which at finite temperatures coexists with the superfluid density that is locked in the condensate. It is interesting to see how our holographic computation relates to this matter.

First, note that in holography it does not seem that the distinction lies in whether or not the charge density is ``normal'' or ``superfluid'', but rather in whether it is inside or outside the horizon, i.e. whether it corresponds to gauge-invariant or deconfined degrees of freedom. These need not be correlated; one could imagine putting some sort of extra charged bulk quanta on top of the superfluid background, thus giving us a ``normal'' component that is nevertheless {\it outside} the horizon. Such an extra contribution would be a gauge-invariant density carried by a charged collective mode of the system, and so would be somewhat analogous to the phonon gas that is studied in conventional Galilean superfluids. This would clearly contribute to the expression \eqref{magnus}; thus in holographic systems we believe one can say cleanly that {\it any} density -- ``normal'' or ``superfluid'' -- contributes in exactly the same way, provided it is outside the horizon. Indeed there is no need to separate them in the calculation. 

Of course in our holographic systems there is a new component to the charge density that does {\it not} contribute: the charge carried by the horizon. This should be roughly thought of as being carried by $N^2$ deconfined gauge-charged degrees of freedom in the dual field theory. There are no analogs of such degrees of freedom in the conventional superfluid; thus in any comparison with conventional systems the term $\sA$ should be set to $0$, and our holographic computation can then be viewed as support for the many-body calculations leading to \eqref{cont}, in which both the normal and superfluid components contribute. 

We stress that one lesson here is that the deconfined holographic superfluid is thus {\it not} a conventional superfluid. This point has been made before: at zero temperature the superfluid geometry is generally Lifshitz, with extra gapless modes that have no place in a traditional superfluid. At finite temperature one is essentially exciting these gapless modes, and we now see that one sharp observable to which they contribute is the coefficient of the Magnus force. 

\subsubsection{Confining superfluids}
Conversely, if one can gap out the gauge theory dynamics (for example by considering it in a confined phase), then one expects to find a conventional superfluid. This problem has been studied in \cite{Nishioka:2009zj,Horowitz:2010jq}, where the confined geometry is given by the AdS soliton \cite{Witten:1998zw}. 

We briefly discuss the construction below. Consider taking a (3+1)-dimensional gauge theory and compactifying one spatial direction on an $S^1$ parametrized by an angle $\psi$. At low energies we are now working with a (2+1) dimensional gauge theory, and depending on the boundary conditions of fields around the $S^1$ we may expect the theory to confine. $\sN = 4$ Super Yang-Mills can provide a precise realization of this \cite{Witten:1998zw}, but we will be working in a bulk gravitational description with extra fields added, and so we will not have a precise CFT dual. The confinement is realized geometrically in that the $S^1$ direction shrinks smoothly to zero size in the interior, ending the geometry at some radial coordinate $r = r^*$. This geometry is the AdS soliton. 

We may now consider adding to this bulk gauge fields and charged scalars to realize a confining superfluid; the phase diagram was mapped out in \cite{Nishioka:2009zj,Horowitz:2010jq}, and there exists a confining superfluid phase extending down to zero temperature over a range of parameters. This has no extra gapless degrees of freedom and so should be dual to a conventional superfluid. It is instructive to see how our arguments above apply to this phase.    

First, it is necessary to identify the ``vortex''. The UV theory is $(3+1)$ dimensional, and so vortex excitations are strings and not points. To obtain a pointlike excitation in the effective low-energy $(2+1)$ dimensional theory, we should arrange for the line to wrap the compact $S^1$. In the bulk the $S^1$ shrinks to zero size; thus the vortex sheet in the (4+1) dimensional bulk should be thought of as a membrane filling the entire $(r,\psi)$ plane at a fixed value of the two remaining spatial coordinates $\vec{x}_0$. Now we have no horizon, and the Berry phase arguments above may be made precise, with $\sA = 0$. 

Similarly, we can repeat the linear response calculation, except with different boundary conditions. At $r = r^*$ we are at the origin of polar coordinates, and so all radial derivatives of bulk fields should be $0$. Thus from \eqref{sigdef} we see that the new boundary condition is simply $\sig_{\pm}(r^*) = 0$. Repeating the calculation we find the same result \eqref{magnus}, except that now there is no dissipative force at all, and $\sA$ is $0$. This is the conventional superfluid result. 

\subsubsection{Speculations at zero temperature}
We now return to the deconfined case. Here it would be very interesting to extend our calculations to zero temperature. We are hindered by the fact that as of yet there has been no explicit construction of the vortex soliton at zero temperature. This may be nontrivial, as generically at zero temperatures one finds Lifshitz solutions; these contain mild ``singularities'' \cite{Kachru:2008yh,Copsey:2010ya,Horowitz:2011gh} which lead local bulk observers to feel diverging tidal forces in the far infrared. It is not obvious to us that the vortex string will maintain its coherence arbitrarily far in the infrared, and the fate of the vortex state thus seems to be an interesting topic for future study. 

If we assume that the soliton remains coherent, then we find the interesting result that at zero temperatures $\sA$ drops to zero by the arguments surrounding \eqref{avanish}. This is then somewhat similar to the situation seen in the electron star, in which we see no anomalous Luttinger count. As pointed out in \cite{arXiv:1111.2606}, however, if we study systems with dilatonic couplings such as \eqref{dilatonic} except with a charged scalar condensate (rather than charged fermions), it seems that there might exist superfluid analogs of the states studied there. In this case some charge remains behind the horizon and coexists with some charge outside stored in the scalar condensate. This system would then exhibit an anomalous Magnus force even at zero temperature.

\section{Conclusion} \label{sec:conc}

In this paper we studied field-theoretical observables that should be thought of as probes of the charge density of the system. For fermionic systems, the observable is the sum of the volumes enclosed by all visible Fermi surfaces. For bosonic systems, the probe is the coefficient of the Magnus force felt by a superfluid vortex. In both cases in ``conventional'' many-body systems, this probe is sensitive to the {\it full} density of the system, as is formalized for fermionic systems by the traditional Luttinger theorem \eqref{lutorig}. In holographic systems, however, we find a different result: for fermionic systems we have a modified Luttinger theorem
\be
\frac{q}{(2\pi)^{d-1}}\sum_i V_i = \jT - \sA, \label{lutrpt} \ee
and for bosonic systems we find
\be
F^i_T = \ep^{ij} \frac{2\pi}{q}(\rho - \sA)v^j \label{sfrpt}
\ee
where $\rho$ is the net $U(1)$ charge density, and $\sA$ is an anomalous term that in the holographic description is the electric field flux sourced by the horizon. This term is not present in conventional systems. Fascinatingly, the horizon provides a mechanism for the violation of the traditional Luttinger theorem and its superfluid analog. Furthermore, the violation takes exactly the same form in both the bosonic and fermionic case. 

We now note several facts that we find interesting regarding this result. First, \HL{from the gravity perspective}, we now have a field-theoretical {\it definition} of the charge carried by the horizon. Given a field theory, simply measure the right-hand side of \eqref{lutrpt} or \eqref{sfrpt} and subtract it from the total charge density $\rho$ to determine $\sA$. Note that this works at finite temperature, even in the spinor case when there are no well-defined Fermi surfaces; the left-hand side is just given by the somewhat messier expression \eqref{finT}, which can in principle be found given a perfect knowledge of the boundary theory spinor Green's function.

Before further interpreting these results, we should discuss the possibility that the term $\sA$ may be an artifact of the gravitational description (or equivalently, an artifact of the large $N$ approximation we are using), and that the true field theory does not have such an anomalous term. It might happen that at finite coupling or $N$ the charged horizons discussed above would be replaced by some object in the gravitational description that would saturate the Luttinger count or contribute to the Magnus force. At this point we should discuss finite-temperature and zero-temperature horizons separately. If we are at finite temperature then the spacetime is completely smooth and we do not expect any such effects to change the results presented here. On the other hand, a zero temperature horizon is usually associated with an infrared subtlety such as a ground-state degeneracy or an IR singularity,
and here it seems quite possible that finite-$N$ or finite-$\lam$ effects could significantly modify the physics. Earlier  in Sec.~\ref{sec:RNBH} we saw such an example:  when the AdS$_2$ dimension of a fermion becomes complex, the horizon of an extremal Reissner-Nordstrom black hole is replaced at finite $N$ by an ``electron star'' which saturates the Luttinger count.




\HL{While this is certainly a logical possibility, from now on we will assume that the effect is real and that an anomalous term $\sA$ truly exists in the dual field theory.} In that case we should attempt to make contact with the many-body systems that are known to exhibit a similar violation \cite{Paramekanti2004,FFL,Z2frac}, in which one finds anomalous terms of precisely the form \eqref{lutrpt} and \eqref{sfrpt} in both the Luttinger count and the Magnus force.\footnote{\HL{Other examples include those in~\cite{Huijse:2011hp}, where the deficit is carried by some gauge {\it variant} spinon Fermi surfaces, which cannot be probed using gauge invariant operators.}}  As mentioned earlier, the existence of the anomalous terms is related to the existence of a deconfined gauge symmetry. 

In fact, in the discussion of~\cite{Paramekanti2004} of fractionalized Fermi liquid and 
superfluid phases, there is a common thread in the derivation of the deficits for the Luttinger count and for the vortex Berry phase. In those derivations one finds relations similar to  \eqref{lutrpt} and \eqref{sfrpt} by considering the system on a torus and threading $2\pi$ flux around one of the cycles while tracking the flow of momentum. Essentially the conventional charged degrees of freedom (e.g. fermionis quasiparticles or superfluid vortices) pick up some momentum under this threading of flux, and if this was the whole story one would find the conventional Luttinger count or Berry phase. However in a fractionalized phase topological excitations of an emergent deconfined gauge theory can also carry some of this momentum, resulting in an anomalous term similar to $\sA$. 

Fascinatingly, in the holographic considerations of this paper, there was also a common thread in the derivations of both results: the observables that we compute are sensitive to bulk {\it charges}, while the actual field theoretical charge density is sensitive instead to the boundary value of the bulk electric field. The relation between these two quantities is provided by Gauss's law: 
\be
\vec{\nabla} \cdot \vec{E}(r) = \rho_{bulk}(r) \ . \label{gauss1}
\ee
However, if we have electric flux emanating from behind a horizon, then there is a component of the boundary theory charge density that has nothing to do with any bulk charges, resulting in a nontrivial $\sA$ that is also clearly related to the existence of deconfined gauge fields. We thus speculate that the same mechanism is at play both in the holographic and condensed matter many-body systems, even if the respective {\it derivations} take rather different routes, and in both cases $\sA$ provides a measure of the deconfined charged degrees of freedom.  It is important to flesh out this connection further.

\HL{We note that it may be possible that the horizon charge degrees of freedom can be probed by some other observables.  Recently it has been proposed in~\cite{Ogawa:2011bz,HiddenFS} 
that the horizon charge degrees of freedom in certain Einstein-Maxwell-dilaton systems  
may reflect ``hidden'' Fermi surfaces which can be probed via entanglement entropy, but not through the correlation functions of gauge invariant operators.}

Further study will be required to understand precisely how such holographic finite-density phases fit into our conventional understanding of quantum matter. One might hope that such study will shed light both on the physics of horizons and of possible phases of matter.

\vspace{0.2in}   \centerline{\bf{Acknowledgements}} \vspace{0.2in} We thank A.~Adams, K.~Balasubramanian, T.~Faulkner, S.~Hartnoll, R.~Loganayagam, V.~Kumar, J.~McGreevy, M.~Roberts, T.~Senthil, B.~Swingle and especially T.~Grover for valuable discussions, and L.~Thompson for very helpful correspondence. HL was supported in part by funds provided by the U.S. Department of Energy (D.O.E.) under cooperative research agreement DE- FG0205ER41360. NI was supported in part by the NSF Grant No. PHY05-51164.

\begin{appendix}
\section{Technical details of spinors} \label{spinorapp}

\subsection{Green's function evaluation of charge density}
Here we perform an alternative derivation of the one-loop charge density that uses an explicit Green's function representation for the spinor density of states. The final result is the same as \eqref{int1} in the main text. We first review some technical aspects of spinor bulk-bulk propagators. Similar considerations can be found in e.g. \cite{Hartman:2010fk}. The Lorentzian spinor action is
\be
S = -\int d^{d+1}x \sqrt{-g} i\bpsi (\Ga^M \sD_M - m) \psi
\ee
By rescaling the spinor the action can be written in momentum space as
\be
S_{\psi} = -\int dr \sqrt{g_{rr}} \frac{d\om}{2\pi}\frac{dk}{(2\pi)^{d-1}} \bPsi(\om,k;r) D \Psi(\om,k;r) \label{bspin2}
\ee
where the bulk differential operator $D$ is
\be
D \equiv i\le(\Ga^{\ur}\p_r + \sqrt{g_{rr}}\le(-m + i \Ga^{\umu}K_{\mu}\sqrt{g^{\mu\mu}}\ri)\ri), \label{Ddef2}
\ee
with $K_t = (-\om - A_t(r)),K_i = k_i$. To define a propagator we will need to invert this operator $D$. $D$ will have eigenfunctions:
\be
D\Psi_n(r,\om,k) = \lam_n \Psi_n(r,\om,k) \label{eigen2}
\ee
To make this equation well-defined we need to specify boundary conditions on the fields both at the horizon and at infinity. At infinity we demand that the spinor fields be normalizable in the usual AdS/CFT sense, as discussed in e.g. \cite{Iqbal:2009fd}. To understand the horizon, let us begin in Euclidean signature, i.e. $\om$ is entirely imaginary. In this case demanding that the equation be regular at the horizon forces the solution to die away there and completely fixes the boundary conditions. 

We now define an inner product between functions as
\be
\langle \Phi, \Psi \rangle \equiv \int dr\;\bPhi \Gamma^{\ur} \Psi 
\ee
$D$ is not {\it quite} hermitian under this inner product; instead, one may show that at complex frequencies $\om$ we have instead 
\begin{align}
\langle \Phi, D_{\om}\Psi \rangle = \langle D_{\om^*}\Phi, \Psi \rangle
\end{align}
Nevertheless, this condition is enough to prove the orthogonality of the eigenfunctions \eqref{eigen2},
\be
\int dr \bPsi_m (\om^*;r)\Ga^{\ur} \Psi_n (\om;r) = \delta_{mn} \label{ortspin} \ .
\ee 
We are not certain how to {\it prove} completeness of the basis \eqref{eigen2}, but if we assume it then the relevant completeness relation is
\be
\sum_n \Psi_n(\om;r) \bPsi_n(\om^*;r')\Gamma^{\ur} = \delta(r-r') \ . \label{comp}
\ee
One can check consistency by integrating both sides of this relation against another eigenfunction $\Psi_p$ and using \eqref{ortspin}. Given these relations, it is easy to see that the bulk-bulk Green's function is
\be
G(\om,k;r,r') = \sum_n \frac{\Psi_n(\om;r) \bPsi_n(\om^*;r')\Ga^{\ur}}{\lam_n(\om)} \label{spineig2}
\ee
By acting on this expression with $D$ and using \eqref{comp} it is simple to verify that $G$ satisfies
\be
D G(\om,k;r,r') = \delta(r - r'),
\ee
which is its defining property. We stress again that $G(\om)$ has been defined by analytic continuation away from Euclidean $\om$, and so in the upper-half plane it will coincide with the retarded propagator, while in the lower half-plane it will coincide with the advanced propagator.

We now use these expressions to rederive \eqref{int1} using a Green's function representation. 

\begin{widetext}
Our starting point is the variation of \eqref{effac}, which yields
\bea
\frac{\delta \Ga[A]}{\delta A_{\tau}(r)} = -\Tr D^{-1}\frac{\delta D}{\delta A_{\tau}(r)} = +iq  \int \frac{d\om_E}{2\pi} \frac{d^{d-1}k}{(2\pi)^{d-1}} \sqrt{g^{\tau\tau}}\sqrt{g_{rr}}\;\tr (\Ga^{\utau}G(i\om_E,k;r,r)),
\eea
where the lowercase trace is over spinor indices. Note that this is essentially the one-loop Feynman diagram shown in Figure \ref{fig:feynman}. This equation, while valid at arbitrary points in the bulk, is somewhat unenlightening; we now insert this into \eqref{varac} and then integrate both sides from the horizon to infinity to find
\be
\frac{1}{g_F^2}\le(\sqrt{-g}F^{rt}\ri)\bigg|^{\infty}_{r_h} = -q\int dr \frac{d\om_E}{2\pi} \frac{d^{d-1}k}{(2\pi)^{d-1}} \sqrt{g^{\tau\tau}}\sqrt{g_{rr}}\;\tr (\Ga^{\utau}G(i\om_E,k;r,r)) \equiv I \label{intanswer}
\ee 

We now need to perform the radial integral to determine $I$ explicitly. We first insert the eigenvalue decomposition \eqref{spineig2} to write the integrand as
\be
I = -q\int dr \frac{d\om_E}{2\pi} \frac{d^{d-1}k}{(2\pi)^{d-1}} \sqrt{g^{\tau\tau}}\sqrt{g_{rr}}\;\tr \le(\Ga^{\utau}\sum_n \frac{\Psi_n(i\om_E;r) \bPsi_n(-i\om_E;r)\Ga^{\ur}}{\lam_n(\om)}\ri) \ . \label{finform}
\ee
We now note that the coupling between the spinor and the gauge field takes the special form $\bPsi(i\om - qA_t)\Psi$. Thus one can essentially trade couplings to the gauge fields for derivatives of the bulk wavefunctions with respect to $\om$; this manipulation is closely related to the Feynman-Hellman theorem. More precisely, we take a derivative with respect to $\om_E$ of the defining equation \eqref{eigen}:
\be
(\p_{\om_E}D) \Psi_n = - (D - \lam_n)\p_{\om_E}\Psi_n + (\p_{\om_E}\lam_n) \Psi_n \label{deriv}
\ee
Now the first term in this expression is
\be
\p_{\om_E}D = -\sqrt{g_{rr}g^{\tau\tau}}\Ga^{\utau}
\ee
which is precisely the coupling leading to the $\Ga^{\utau}$ in \eqref{finform}. Thus we may insert \eqref{deriv} into \eqref{finform} and then integrate by parts to transfer the $D - \lam_n$ to the other wavefunction factor, annihilating it. We then find only 
\be
I = iq\int\frac{d\om_E}{2\pi} \frac{d^{d-1}k}{(2\pi)^{d-1}} \sum_n \frac{\p_{\om_E}\lam_n}{\lam_n} \int dr \bPsi_n(-i\om_E;r)\Ga^{\ur}\Psi_n(i\om_E,r)
\ee
\end{widetext}
The radial integral is now done by appealing to the orthogonality relation \eqref{ortspin}. We find finally 
\be
I =  -iq \int \frac{d\om_E}{(2\pi)}\frac{d^{d-1}k}{(2\pi)^{d-1}}  \sum_n \frac{\p}{\p\om_E}\log \lam_n(\om_E), 
\ee
which is \eqref{int1} in the text. 
\subsection{Finite temperature Luttinger theorem}
We now present the manipulations involved in deriving a finite-temperature ``Luttinger theorem'', i.e. an expression relating the charge density outside a finite temperature horizon to a property of the spinor Green's function. We begin with the expression \eqref{int1}, except at finite temperature, so that the integral over frequencies becomes a Matsubara sum:
\be
I = iq T \sum_{i\om_n} \int \frac{d^{d-1} k}{(2\pi)^{d-1}} \sum_n \frac{\p}{\p\om_E}\log \lam_n
\ee \label{intT}
where the fermionic Matsubara frequencies are
\be
\om_n = (2n+1)\pi T \qquad n \in \mathbb{Z}
\ee
We now perform some standard finite-temperature field-theoretical manipulations. We first write the Matsubara sum as a contour integral:
\be
\sum_{i\om_n} \to -\frac{1}{2\pi i}\oint d\om f(\om) \label{Tcont}
\ee
where the function $f(\om)$ has poles at all the Matsubara frequencies, and so is
\be
f(\om) = \frac{\beta}{\exp(\beta \om) + 1} = -\frac{\beta}{2}\le(\tanh\le(\frac{\beta\om}{2}\ri) - 1\ri)
\ee
The relevant contour is shown in Figure 

\begin{figure}[h]
\begin{center}
\includegraphics[scale=0.50]{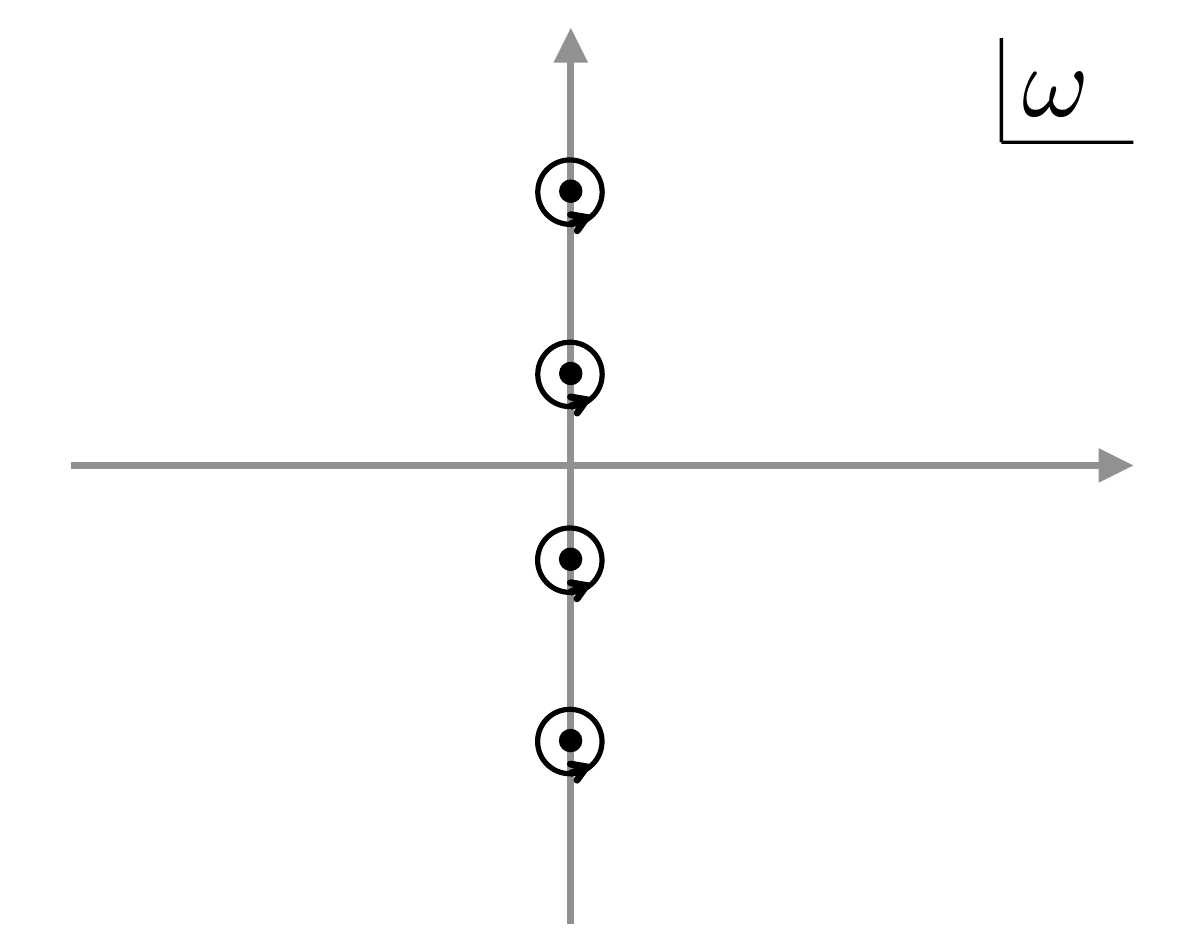}
\end{center}
\vskip -0.5cm
\caption{Integration contour used in \eqref{Tcont}.}
\label{fig:wiggly}
\end{figure}

Putting this in we find the formula
\be
I = -\frac{iq T}{2\pi} \int \frac{d^{d-1} k}{(2\pi)^{d-1}} \oint d\om f(\om) \sum_n \frac{\p}{\p\om} \log \lam_n(\om)
\ee
Now we use the identity $\pi\tanh(\pi x) = i\le(\psi\le(\ha + i x\ri) - \psi\le(\ha - i x\ri)\ri)$ to rewrite the hyperbolic tangent and get the following representation of the Fermi distribution:
\be
f(\om) = -\frac{\beta}{2}\le(\frac{i}{\pi}\le(\psi\le(\ha + \frac{i\beta\om}{2\pi}\ri) - \psi\le(\ha - \frac{i\beta\om}{2\pi}\ri)\ri)-1\ri)
\ee
This is useful because the digamma function $\psi$ only has poles at negative integer values of its argument; thus the first digamma has poles only in the upper-half plane, and the second has them only in the lower-half-plane. So we now do each digamma integral separately. 

For the first one, we deform the contour downwards into the lower-half plane, as in Figure \ref{fig:Tman}. We are now working with the {\it analytic continuation} of $\lam_n(\om)$ to the lower-half plane; this has a collection of zeros at complex frequencies $\om = \Om_a(k)$, each of which is a quasinormal mode. The integral over $\om$ will pick up a contribution times $(-2\pi i)$ from each of these zeros\footnote{The minus sign is because the orientation of the integral is backwards in the lower-half plane}. We do the same thing for the other digamma, except that we now deform the contour upwards, picking up a contribution from $\Om_a^*(k)$.  

We find once the dust settles
\be 
I = -q\int \frac{d^{d-1}k}{(2\pi)^{d-1}} \sum_a \le(\ha - \frac{1}{\pi} \Im \psi\le(\ha + \frac{i\beta \Om_a(k)}{2\pi}\ri)\ri)
\ee
This is \eqref{finT} in the main text: its interpretation is discussed there.  

\begin{widetext}
\begin{center}
\begin{figure}[h]
\includegraphics[scale=0.50]{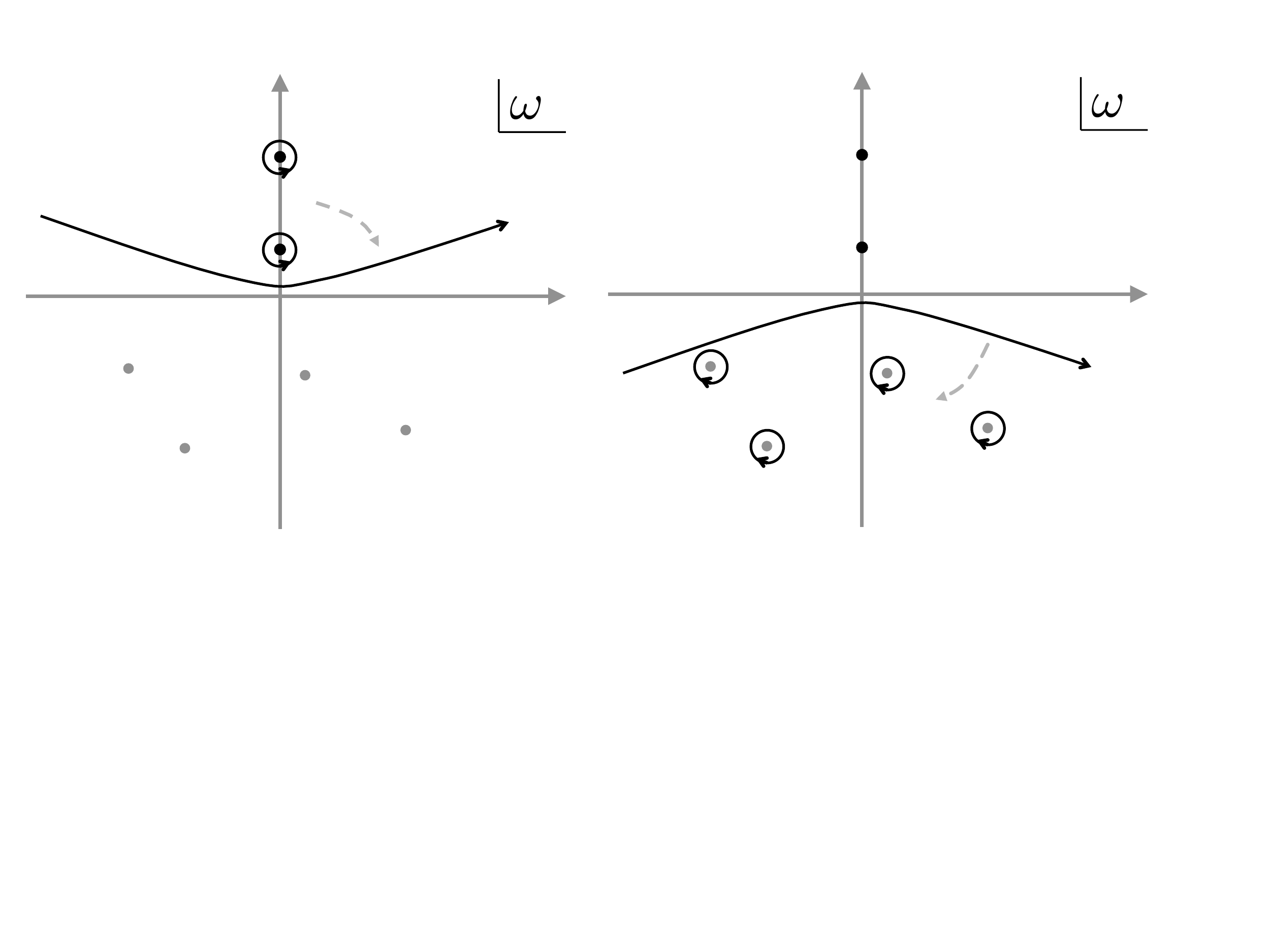}

\vskip -0.5cm
\caption{Contour manipulations discussed above, for digamma function with poles in the upper-half-plane. Gray circles indicate zeros of $\lam^I(\om)$, which are present if $\lam(\om)$ is analytically continued into the lower-half-plane. Note a similar manipulation is done on the other digamma function.}
\label{fig:Tman}
\end{figure}
\end{center}
\end{widetext}

\section{Technical details of vortices} \label{scalarapp}
Here we review some details about boundary conditions on the vortex soliton that were glossed over in the main text; some of these considerations can also be found in \cite{Keranen:2009re}. Recall that the bulk action is
\be
S = -\int d^{4}x \sqrt{-g}\le( \frac{1}{4g_F^2} F^2 + (D\phi)^{\dagger} D\phi + V(\phi^{\dagger}\phi)\ri) \label{scalarac2}
\ee
We will use $r$ for the bulk holographic direction and will parametrize the field theory spatial directions by polar coordinates $(\rho,\th)$.
 
In the superfluid phase the condensate $\phi(r)$ has a profile; it is large in the interior and dies away at infinity so that it represents a normalizable state in the theory. In the bulk the symmetry is gauged by $A$; thus there is some sense where we have a superconductor in the deep interior that is crossing over into an unHiggsed phase near the boundary. If we now want to construct a vortex on top of this background superfluid, we put an Abrikosov-Nielsen-Olesen string on top of this background, as discussed around \eqref{soliton}. The scalar profile takes the form
\be
\phi(r) = v(r)e^{i\al(\th)} \qquad \al(\th) = \th \ .
\ee

Now this string carries magnetic flux $F^{\rho\th}$ in the bulk, but if we are interested in looking at a normalizable state then this magnetic flux must not penetrate all the way to the boundary -- if it did, it would then define a field-theory state with an applied magnetic field, which is not what we are looking for. Thus something must happen to the magnetic field. It is clear that it must spread out along the field theory directions, and thus turn from a radial magnetic field into a field parallel to the boundary, i.e. it turns into a component of $F^{r\th}$ rather than $F^{\rho\th}$. See Figure \ref{fig:vortices}. However this component of the bulk field strength tensor can now be interpreted as a circular current flow in the boundary, i.e.
\be
\langle j^{\th} \rangle_{QFT} = -\frac{1}{g_F^2}\sqrt{-g}F^{r\th}(r \to \infty) \label{jthdef}
\ee
To demonstrate this, let us compute this current far from the vortex core. The dependence on $\rho$ can be neglected, and the relevant Maxwell equation following from \eqref{scalarac2} is
\be
\frac{1}{g_F^2} \p_r \le(\sqrt{-g}F^{r\th}\ri) + 2v^2 q g^{\th\th}\sqrt{-g}\le(\p_{\th}\al - qA_{\th}\ri) = 0 \label{ode}
\ee
This is essentially a forced equation for $A_{\th}$, where the forcing term is given by $\p_{\th}\al = 1$. We may easily solve it by setting, 
\be
A_{\th}(r) = \frac{1}{q}(1 - a_{\th}(r))
\ee
where $a_{\th}(r)$ obeys the homogenous part of the same equation \eqref{ode}, i.e. 
\be
\frac{1}{g_F^2 }\p_r\le(\sqrt{-g}g^{rr}g^{\th\th} \p_{r} a_{\th}\ri) - 2v^2 q g^{\th\th}\sqrt{-g} q a_{\th} = 0 \ . \label{ode2}
\ee
Furthermore, we require that $a_{\th}(r)$ be regular at the horizon and satisfy the boundary condition $a_{\th}(r\to\infty) = 1$, such that at infinity $A_{\th}$ (the field theory source) vanishes. From \eqref{jthdef} $j^{\th}$ is simply given by $-\frac{1}{q}$ times the radial derivative of $a_{\th}$ at infinity. 

Now note that the equation obeyed by $a_{\th}(r)$ is simply that for a spatially homogenous gauge fluctuation on the superfluid background; the presence of the vortex cannot be seen in \eqref{ode2}. In fact, since $a_{\th}$ is regular at the horizon, \eqref{ode2} is precisely the equation that we solve if we want to compute the two point function of the current $G_{\th\th}$ at zero frequency and momentum. It is shown in \cite{Iqbal:2008by} that on such a field configuration the boundary theory Green's function is precisely the ratio of the radial derivative $\p_{r}a_{\th}$ to $a_{\th}$ at infinity; converting from $G_{\th\th}$ to the cartesian $G_{xx}$ we conclude that the circular current $j^{\th}$ satisfies
\be
j_{\th}(\rho) = \frac{G_{xx}(\om = 0, k = 0)}{q}
\ee
Now in a relativistic superfluid one way to define the superfluid density $\rho_s$ is via the zero-frequency value of $G_{xx}$, e.g. \cite{hep-ph/0011246,arXiv:0809.4870,arXiv:0906.4810}:
\be
G_{xx}(\om = 0, k = 0) = \frac{\rho_s}{\mu}
\ee
We thus conclude that the circulation around the vortex is
\be
\oint d\th j_{\th} = \frac{2\pi}{q} \frac{\rho_s}{\mu}, \label{circ}
\ee
as is expected.

Now in the calculation in the main text we assume that all of the magnetic flux remains locked in the string all the way up to infinity, and thus we are studying the field theory with some magnetic flux inserted into the system. We should imagine this as being a pinning potential for the vortex: the location of this flux defines the position of the vortex $\vec{x_0}$ in the field theory, and presumably the force calculated in \eqref{magnus} actually acts on the apparatus that is producing this flux. In a truly normalizable vortex state this flux would not be there, the field lines would spread out parallel to the boundary, and the vortex would be free to move around. However in this situation it is somewhat difficult to precisely define the force on the vortex; to extract it we would need to examine the equation of motion of a dynamical vortex, which appears to be a somewhat more difficult problem then the one solved in the text. If such a calculation could be done we expect the answer to be roughly the same, as the only distinction between these cases is near the boundary where the scalar profile (and thus the associated charge density) is small.

\begin{widetext}
\begin{center}
\begin{figure}[h]
\includegraphics[scale=0.70]{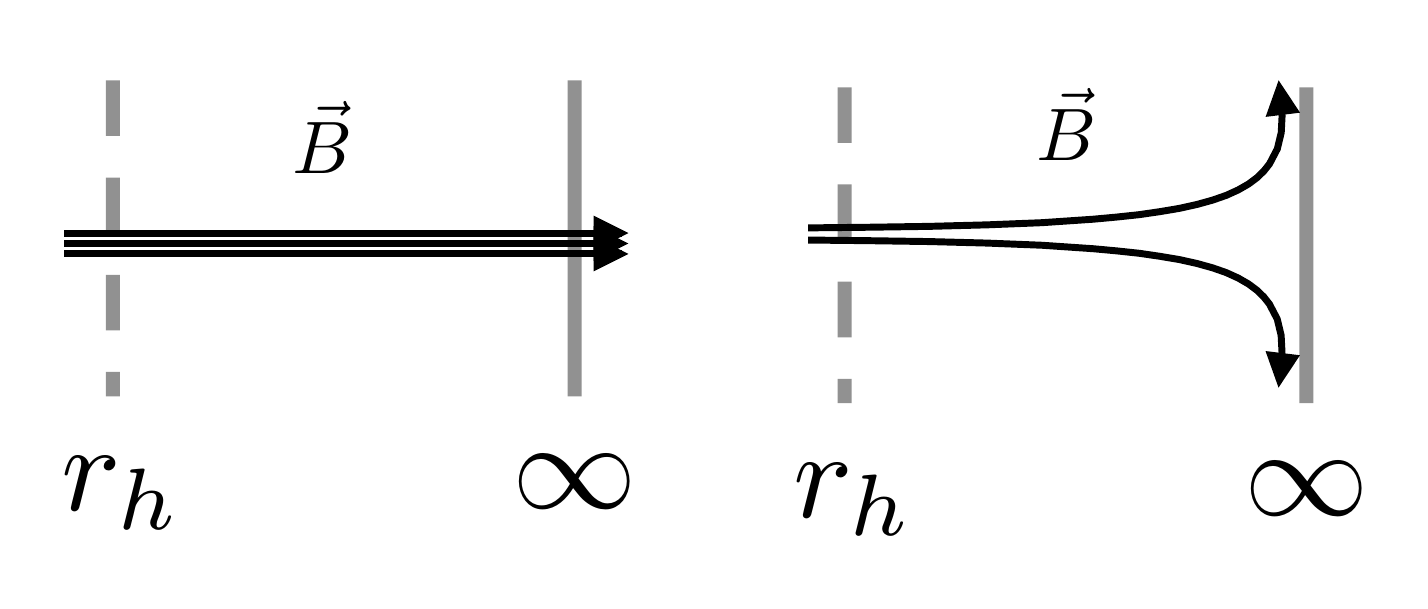}
\vskip -0.5cm
\caption{Two possible vortex solutions. {\it Left}: the vortex has magnetic flux enclosed in the string all the way to the boundary. This is the vortex that we study in the bulk of the paper, but note that it is not a normalizable solution as it involves a nontrivial $F_{xy}$ source in the field theory. The location of this flux defines the position of the vortex, and the vortex is not free to move. {\it Right}: the vortex solution analyzed in this appendix. This has no magnetic field perpendicular to the boundary; the field tangential to the boundary can be interpreted as circulation, as in \eqref{circ}. This vortex is a truly normalizable solution, and is free to move.}
\label{fig:vortices}
\end{figure}
\end{center}
\end{widetext}

\end{appendix}


\begin{thebibliography}{9}
%

\bibitem{Huijse:2011hp}
  L.~Huijse and S.~Sachdev,
  ``Fermi surfaces and gauge-gravity duality,''
  Phys.\ Rev.\  D {\bf 84}, 026001 (2011)
  [arXiv:1104.5022 [hep-th]].
  

 
  \bibitem{Luttinger}
  J.~M.~Luttinger, ``Fermi Surface and Some Simple Equilibrium Properties of a System of Interacting Fermions,'' Phys.\ Rev.\ {\bf 119}, 1153 (1960). 
  
   \bibitem{Oshikawa}
  M.~Oshikawa, ``Topological Approach to Luttinger's Theorem and the Fermi Surface of a Kondo Lattice,'' Phys.\ Rev.\ Lett. {\bf 84}, 3370 (2000) [arXiv:cond-mat/0002392].
  
    
 \bibitem{Paramekanti2004}
  A.~Paramekanti and A.~Vishwanath, ``Extending Luttinger's theorem to $Z_2$ fractionalized phases of matter,'' Phys.~Rev.~B {\bf 70}, 245118 (2004) [arXiv:cond-mat/0406619].
  
  \bibitem{FFL}
  T.~Senthil, S.~Sachdev, and M.~Vojta, ``Fractionalized Fermi Liquids,'' Phys.~Rev.~Lett. {\bf 90}, 216403 (2003) [arXiv:cond-mat/0209144].
  
  \bibitem{Z2frac}
  T.~Senthil and M.~P.~A.~Fisher, ``$Z_2$ gauge theory of electron fractionalization in strongly correlated systems,'' Phys.~Rev.~B {\bf 62}, 7850 (2000) [arXiv:cond-mat/9910224].
  
  \bibitem{AdS/CFT}
J.~M.~Maldacena,
``The large N limit of superconformal field theories and supergravity,''
Adv.\ Theor.\ Math.\ Phys.\  {\bf 2}, 231 (1998) [arXiv:hep-th/9711200];
 E.~Witten,
  ``Anti-de Sitter space and holography,''
{\it ibid.} 253 (1998) [arXiv:hep-th/9802150];

S.~S.~Gubser, I.~R.~Klebanov and A.~M.~Polyakov,
``Gauge theory correlators from non-critical string theory,''
Phys.\ Lett.\ B {\bf 428}, 105 (1998) [arXiv:hep-th/9802109].


  
  \bibitem{RNBH}
  L.~J.~Romans, ``Supersymmetric, cold and lukewarm black holes in cosmological Einstein-Maxwell theory,''
  Nucl.\ Phys.\  B {\bf 383}, 395 (1992)
  [arXiv:hep-th/9203018];
  A.~Chamblin, R.~Emparan, C.~V.~Johnson and R.~C.~Myers, ``Charged AdS black holes and catastrophic holography,''
  Phys.\ Rev.\  D {\bf 60}, 064018 (1999)
  [arXiv:hep-th/9902170].

\bibitem{Goldstein:2009cv}
  K.~Goldstein, S.~Kachru, S.~Prakash, S.~P.~Trivedi,
  ``Holography of Charged Dilaton Black Holes,''
  JHEP {\bf 1008}, 078 (2010).
  [arXiv:0911.3586 [hep-th]].
  
\bibitem{Charmousis:2010zz} 
  C.~Charmousis, B.~Gouteraux, B.~S.~Kim, E.~Kiritsis and R.~Meyer,
  ``Effective Holographic Theories for low-temperature condensed matter systems,''
  JHEP {\bf 1011}, 151 (2010)
  [arXiv:1005.4690 [hep-th]].
  
\bibitem{Gubser:2009qt} 
  S.~S.~Gubser and F.~D.~Rocha,
  ``Peculiar properties of a charged dilatonic black hole in $AdS_5$,''
  Phys.\ Rev.\ D {\bf 81}, 046001 (2010)
  [arXiv:0911.2898 [hep-th]].
  
  
    
\bibitem{Hartnoll:2009ns}
  S.~A.~Hartnoll, J.~Polchinski, E.~Silverstein and D.~Tong,
  ``Towards strange metallic holography,''
  JHEP {\bf 1004}, 120 (2010)
  [arXiv:0912.1061 [hep-th]].
  
   \bibitem{Hartnoll:2010gu}
  S.~A.~Hartnoll, A.~Tavanfar,
  ``Electron stars for holographic metallic criticality,''
  Phys.\ Rev.\  {\bf D83}, 046003 (2011).
  [arXiv:1008.2828 [hep-th]].

\bibitem{Arsiwalla:2010bt}
  X.~Arsiwalla, J.~de Boer, K.~Papadodimas and E.~Verlinde, ``Degenerate Stars and Gravitational Collapse in AdS/CFT,''
  JHEP {\bf 1101}, 144 (2011)
  [arXiv:1010.5784 [hep-th]].
  
\bibitem{vCubrovic:2010bf}
  M.~Cubrovic, J.~Zaanen and K.~Schalm, ``Constructing the AdS dual of a Fermi liquid: AdS Black holes with Dirac hair,''
  arXiv:1012.5681 [hep-th].
  
\bibitem{Harada:2011aa} 
  M.~Harada, S.~Nakamura and S.~Takemoto,
  ``Holographic Mean-Field Theory for Baryon Many-Body Systems,''
  arXiv:1112.2114 [hep-th].
    
  
    \bibitem{holographicsc}
  S.~S.~Gubser, ``Breaking an Abelian gauge symmetry near a black hole horizon,''
  Phys.\ Rev.\  D {\bf 78}, 065034 (2008)
  [arXiv:0801.2977 [hep-th]];
  S.~A.~Hartnoll, C.~P.~Herzog and G.~T.~Horowitz, ``Building a Holographic Superconductor,''
  Phys.\ Rev.\ Lett.\  {\bf 101}, 031601 (2008)
  [arXiv:0803.3295 [hep-th]].
  
  

  \bibitem{arXiv:1111.2606} 
  S.~A.~Hartnoll and L.~Huijse,
  ``Fractionalization of holographic Fermi surfaces,''
  arXiv:1111.2606 [hep-th].
  
  \bibitem{arXiv:1006.3794} 
  S.~Sachdev,
  ``Holographic metals and the fractionalized Fermi liquid,''
  Phys.\ Rev.\ Lett.\ \ {\bf 105}, 151602  (2010)
  [arXiv:1006.3794 [hep-th]].  
  
\bibitem{arXiv:1012.0299} 
  S.~Sachdev,
``The landscape of the Hubbard model,''
  arXiv:1012.0299 [hep-th].
  
   \bibitem{arXiv:1106.4324} 
  S.~A.~Hartnoll,
  ``Horizons, holography and condensed matter,''
  arXiv:1106.4324 [hep-th].
  
    
    \bibitem{Hartnoll:2011dm}
  S.~A.~Hartnoll, D.~M.~Hofman, D.~Vegh,
  ``Stellar spectroscopy: Fermions and holographic Lifshitz criticality,''
  JHEP {\bf 1108}, 096 (2011).
  [arXiv:1105.3197 [hep-th]].
  
  \bibitem{Iqbal:2011in}
  N.~Iqbal, H.~Liu, M.~Mezei,
  ``Semi-local quantum liquids,''
  arXiv:1105.4621 [hep-th].
  

  
\bibitem{Sachdev:2011ze}
  S.~Sachdev,
  ``A model of a Fermi liquid using gauge-gravity duality,''
  Phys.\ Rev.\  D {\bf 84}, 066009 (2011)
  [arXiv:1107.5321 [hep-th]].
 
  
      
\bibitem{SachdevNew}
S.~Sachdev, ``What can gauge-gravity duality teach us about condensed matter physics?'' arXiv:1108.1197 [cond-mat.str-el]. 
   
\bibitem{Witten:1998zw}
  E.~Witten,
  ``Anti-de Sitter space, thermal phase transition, and confinement in gauge theories,''
  Adv.\ Theor.\ Math.\ Phys.\  {\bf 2}, 505-532 (1998).
  [hep-th/9803131].
  
   
   
\bibitem{Liu:2009dm}
  H.~Liu, J.~McGreevy, and D.~Vegh, ``Non-Fermi liquids from holography,'' arXiv:0903.2477 [hep-th]

\bibitem{Lee:2008xf}  
  S.-S. Lee, ``A Non-Fermi Liquid from a Charged Black Hole; A Critical Fermi Ball,'' Phys.~Rev.~D \textbf{79}, 086006 (2009) [arXiv: 0809.3402 [hep-th]]

\bibitem{Cubrovic:2009ye}  
  M.~Cubrovic, J.~Zaanen, and K.~Schalm, ``String Theory, Quantum Phase Transitions and the Emergent Fermi-Liquid,'' Science \textbf{325}, 439--444 (2009) [arXiv:0904.1993 [hep-th]].
  
\bibitem{Faulkner:2009wj}
  T.~Faulkner, H.~Liu, J.~McGreevy and D.~Vegh,
  ``Emergent quantum criticality, Fermi surfaces, and AdS2,''
  arXiv:0907.2694 [hep-th].

\bibitem{Iqbal:2011ae} 
  N.~Iqbal, H.~Liu and M.~Mezei,
  ``Lectures on holographic non-Fermi liquids and quantum phase transitions,''
  arXiv:1110.3814 [hep-th].

\bibitem{Iqbal:2009fd}
  N.~Iqbal and H.~Liu,
  ``Real-time response in AdS/CFT with application to spinors,''
  Fortsch.\ Phys.\  {\bf 57}, 367 (2009)
  [arXiv:0903.2596 [hep-th]].
  
\bibitem{Denef:2009kn}
  F.~Denef, S.~A.~Hartnoll, S.~Sachdev,
  ``Black hole determinants and quasinormal modes,''
  Class.\ Quant.\ Grav.\  {\bf 27}, 125001 (2010).
  [arXiv:0908.2657 [hep-th]].

\bibitem{Denef:2009yy}
  F.~Denef, S.~A.~Hartnoll, S.~Sachdev,
  ``Quantum oscillations and black hole ringing,''
  Phys.\ Rev.\  {\bf D80}, 126016 (2009).
  [arXiv:0908.1788 [hep-th]].
  
  \bibitem{Hartnoll:2009sz} 
  S.~A.~Hartnoll,
  ``Lectures on holographic methods for condensed matter physics,''
  Class.\ Quant.\ Grav.\  {\bf 26}, 224002 (2009)
  [arXiv:0903.3246 [hep-th]].
  
    \bibitem{Herzog:2009xv} 
  C.~P.~Herzog,
  ``Lectures on Holographic Superfluidity and Superconductivity,''
  J.\ Phys.\ A A {\bf 42}, 343001 (2009)
  [arXiv:0904.1975 [hep-th]].
  
  
  \bibitem{Horowitz:2010gk} 
  G.~T.~Horowitz,
  ``Introduction to Holographic Superconductors,''
  arXiv:1002.1722 [hep-th].
  
  
  \bibitem{Haldane1985} 
  F.~D.~M.~Haldane and Y-S.~Wu, ``Quantum Dynamics and Statistics of Vortices in Two-Dimensional Superfluids,'' Phys.~Rev.~Lett. {\bf 55}, 2887 (1985).
  \bibitem{Ao1993} P.~Ao and D.~Thouless, ``Berry's Phase and the Magnus Force for a Vortex Line in a Superconductor,'' Phys.~Rev. Lett.~ {\bf 70}, 2158 (1993)
  
  \bibitem{controversy}E.~B.~Sonin, ``The Magnus force in superfluids and superconductors,'' Phys.\ Rev.\ B {\bf 55} 485 (1997) [arXiv:cond-mat/9606099]; C.~Wexler, ``Magnus and Iordanakii Forces in Superfluids,'' Phys.\ Rev.\ Lett. {\bf 79} 1321 (1997) [arXiv:cond-mat/961211];  D.~J.~Thouless, P.~Ao, and Q.~Niu ``Transverse force on a quantized vortex in a superfluid,'' Phys.\ Rev.\ Lett. {\bf 76} 3758 (1996) [arXiv:cond-mat/960319].
  \bibitem{lara}L.~Thompson, ``Equation of motion of a quantum vortex,'' Ph.D thesis, University of British Columbia. Available online at {\tt https://circle.ubc.ca/handle/2429/30460.}
  
  \bibitem{hep-ph/0011246} 
  D.~T.~Son,
  ``Hydrodynamics of relativistic systems with broken continuous symmetries,''
  Int.\ J.\ Mod.\ Phys.\ A\ {\bf 16S1C}, 1284  (2001)
  [hep-ph/0011246].
  
  \bibitem{arXiv:0809.4870} 
  C.~P.~Herzog, P.~K.~Kovtun and D.~T.~Son,
  ``Holographic model of superfluidity,''
  Phys.\ Rev.\ D\ {\bf 79}, 066002  (2009)
  [arXiv:0809.4870 [hep-th]].
  
  \bibitem{arXiv:0906.4810} 
  C.~P.~Herzog and A.~Yarom,
  ``Sound modes in holographic superfluids,''
  Phys.\ Rev.\ D\ {\bf 80}, 106002  (2009)
  [arXiv:0906.4810 [hep-th]].
  
  \bibitem{Albash:2009iq}
  T.~Albash and C.~V.~Johnson,
  ``Vortex and Droplet Engineering in Holographic Superconductors,''
  Phys.\ Rev.\  D {\bf 80}, 126009 (2009)
  [arXiv:0906.1795 [hep-th]].
  
  \bibitem{Montull:2009fe}
  M.~Montull, A.~Pomarol, P.~J.~Silva,
  ``The Holographic Superconductor Vortex,''
  Phys.\ Rev.\ Lett.\  {\bf 103}, 091601 (2009).
  [arXiv:0906.2396 [hep-th]].
  
\bibitem{Keranen:2009re}
  V.~Keranen, E.~Keski-Vakkuri, S.~Nowling, K.~P.~Yogendran,
  ``Inhomogeneous Structures in Holographic Superfluids: II. Vortices,''
  Phys.\ Rev.\  {\bf D81}, 126012 (2010).
  [arXiv:0912.4280 [hep-th]].
  
  \bibitem{Maeda:2009vf} 
  K.~Maeda, M.~Natsuume and T.~Okamura,
 ``Vortex lattice for a holographic superconductor,''
  Phys.\ Rev.\ D {\bf 81}, 026002 (2010)
  [arXiv:0910.4475 [hep-th]].
  
  \bibitem{coleman}
S.~Coleman, ``Aspects of Symmetry,'' Cambridge University Press, New York 1985.

\bibitem{Herzog:2006gh}
  C.~P.~Herzog, A.~Karch, P.~Kovtun, C.~Kozcaz, L.~G.~Yaffe,
  ``Energy loss of a heavy quark moving through N=4 supersymmetric Yang-Mills plasma,''
  JHEP {\bf 0607}, 013 (2006).
  [hep-th/0605158].
  
  \bibitem{Gubser:2006bz}
  S.~S.~Gubser,
  ``Drag force in AdS/CFT,''
  Phys.\ Rev.\  D {\bf 74}, 126005 (2006)
  [arXiv:hep-th/0605182].

\bibitem{Iqbal:2008by}
  N.~Iqbal and H.~Liu,
  ``Universality of the hydrodynamic limit in AdS/CFT and the membrane
  paradigm,''
  Phys.\ Rev.\  D {\bf 79}, 025023 (2009)
  [arXiv:0809.3808 [hep-th]].


\bibitem{Nishioka:2009zj}
  T.~Nishioka, S.~Ryu and T.~Takayanagi,
  ``Holographic Superconductor/Insulator Transition at Zero Temperature,''
  JHEP {\bf 1003}, 131 (2010)
  [arXiv:0911.0962 [hep-th]].
  
\bibitem{Horowitz:2010jq}
  G.~T.~Horowitz, B.~Way,
  ``Complete Phase Diagrams for a Holographic Superconductor/Insulator System,''
  JHEP {\bf 1011}, 011 (2010).
  [arXiv:1007.3714 [hep-th]].
  
  \bibitem{Kachru:2008yh}
  S.~Kachru, X.~Liu, M.~Mulligan,
  ``Gravity Duals of Lifshitz-like Fixed Points,''
  Phys.\ Rev.\  {\bf D78}, 106005 (2008).
  [arXiv:0808.1725 [hep-th]].
  
  \bibitem{Copsey:2010ya}
  K.~Copsey and R.~Mann,
 ``Pathologies in Asymptotically Lifshitz Spacetimes,''
  JHEP {\bf 1103}, 039 (2011)
  [arXiv:1011.3502 [hep-th]].
  
\bibitem{Horowitz:2011gh}
  G.~T.~Horowitz and B.~Way,
  ``Lifshitz Singularities,''
  arXiv:1111.1243 [hep-th].

\bibitem{Ogawa:2011bz} 
  N.~Ogawa, T.~Takayanagi and T.~Ugajin,
  ``Holographic Fermi Surfaces and Entanglement Entropy,''
  arXiv:1111.1023 [hep-th].
  
  \bibitem{HiddenFS}
L.~Huijse, S.~Sachdev, B.~Swingle, ``Hidden Fermi surfaces in compressible states of gauge-gravity duality,'' arXiv:1112.0573 [cond-mat.str-el].

  
\bibitem{Hartman:2010fk}
  T.~Hartman, S.~A.~Hartnoll,
  ``Cooper pairing near charged black holes,''
  JHEP {\bf 1006}, 005 (2010).
  [arXiv:1003.1918 [hep-th]].
  

\end{thebibliography}
\end{document}